\documentclass[twocolumn,final]{pasj01}

\usepackage{url}
\usepackage{lscape}
\usepackage{ulem}

\Received{}
\Accepted{}
 
\usepackage{mathpazo}
\usepackage[T1]{fontenc} 
\newcommand\ZAMS{{\rm ZAMS}}

\newcommand\co{{\rm co}}

\newcommand\HMXB{{\rm HMXB}}
\newcommand\LMXB{{\rm LMXB}}
\newcommand\galaxy{{\rm Gal}}
\newcommand\ellE{\ell_{\rm Edd}}

\newcommand\LEdd{L_{\rm Edd}}
\newcommand{\heii}{He~{\sc ii}}
\graphicspath{{./}{figures/}}

\begin{document} 

\title{A Fundamental Plane in X-ray Binary Activity of External Galaxies}

\author{Yoshiyuki \textsc{Inoue}\altaffilmark{1,2,3}}
\altaffiltext{1}{Department of Earth and Space Science, Graduate School of Science, Osaka University, Toyonaka, Osaka 560-0043, Japan}
\altaffiltext{2}{Interdisciplinary Theoretical \& Mathematical Science Program (iTHEMS), RIKEN, 2-1 Hirosawa, Saitama 351-0198, Japan}
\altaffiltext{3}{Kavli Institute for the Physics and Mathematics of the Universe (WPI), UTIAS, The University of Tokyo, Kashiwa, Chiba 277-8583, Japan}
\email{yinoue@astro-osaka.jp}

\author{Kiyoto \textsc{Yabe},\altaffilmark{3}}
\email{kiyoto.yabe@ipmu.jp}

\author{Yoshihiro \textsc{Ueda}\altaffilmark{4}}
\altaffiltext{4}{Department of Astronomy, Kyoto University, Kyoto 606-8502, Japan}
\email{ueda@kusastro.kyoto-u.ac.jp}

\KeyWords{X-rays: binaries --- binaries: general --- stars: black holes}

\maketitle

\begin{abstract}
We construct a new catalog of extragalactic X-ray binaries (XRBs) {by matching} the latest {{\it Chandra}} source {catalog with local galaxy catalogs}. Our XRB catalog contains {4430} XRBs hosted by {237} galaxies within $\sim130$~Mpc. As XRBs dominate the X-ray activity in galaxies, the catalog enables us to study the correlations between the total X-ray luminosity of a galaxy $L_{X,\rm tot}$, star formation rate $\dot{\rho}_\star$, and stellar mass $M_\star$. As previously reported, $L_{X,\rm tot}$ is correlated with $\dot{\rho}_\star$ and $M_\star$. In particular, we find that there is a fundamental plane in those three parameters as $\log L_{X,\rm tot}={38.80^{+0.09}_{-0.12}}+\log(\dot{\rho}_\star + \alpha M_\star)$, where $\alpha = {(3.36\pm1.40)\times10^{-11}}\ {\rm yr^{-1}}$. In order to investigate this relation, we construct a phenomenological binary population synthesis model. We find that the high mass XRB and low mass XRB fraction in formed compact object binary systems is $\sim 9$\% and $\sim0.04$\%, respectively. Utilizing the latest XMM-Newton, and Swift X-ray source catalog data sets, additional XRB candidates are also found resulting in 5757 XRBs hosted by 311 galaxies.
\end{abstract}

\section{Introduction} 
\label{sec:intro}
Recent discoveries of gravitational waves from merging black holes (BHs) and neutron stars (NSs) have finally opened the multi-messenger astronomy era  (\cite{Abbott2016_BBH, Abbott2017_BNS}). However, the nature of those binary systems is not fully understood yet. In order to elucidate the formation mechanism of those binary systems, various binary population synthesis models are proposed in the literature (e.g., \cite{Belczynski2008, Belczynski2010, Belczynski2016_Nat, Belczynski2016,  Mandel2016, Pavlovskii2017, Marchant2017, Kruckow2018, Mapelli2018}). 

Here, X-ray binaries (XRBs) can provide an independent test of those binary population synthesis models. XRBs are close binary systems found in nearby galaxies and radiate most of their emission in the X-ray band (see, e.g., \cite{Done2007}). They are powered by mass accretion from a companion star onto a compact object (i.e., BH and NS) either through Roche lobe overflow or stellar-wind fed process. After the bright X-ray phase, XRBs evolve into various compact binary systems such as {gravitational wave} merger sources and millisecond pulsars. 

Among XRBs, the formation of luminous XRBs, so-called ultra-luminous X-ray sources (ULXs; see \cite{Kaaret2017} for recent reviews) would be tightly related to the coalescences of currently observed binary BHs \citep{Abbott2016_BBH, Nitz2019, Abbott2020arXiv201014527A}. Most of the observed binary BHs are found to be composed of BHs with $\gtrsim20M_\odot$ \citep{Abbott2016_BBH, Nitz2019}, more massive than known Galactic stellar BHs $\approx10M_\odot$ \citep{Casares2014}\footnote{At the Galactic center, there is a central supermassive BHs, e.g., Sgr A* \citep{Boehle2016}. Recently, intermediate-mass BHs are also reported inside the Galaxy (e.g., \cite{Oka2017, Liu2019}). However, these have not been firmly confirmed yet (see, e.g., \cite{ElBadry2020} for LB-1).}. Binary population synthesis models predict that the common envelope scenario \citep{Pavlovskii2017} and the chemically homogeneous evolution scenario \citep{Marchant2017} could form those binary BHs through ULX phases. It is also known that the local ULX formation rate is consistent with the measured binary BH merger rate \citep{Inoue2016, Finke2017}. Therefore, statistical information of XRBs, including ULXs, would be keys to exploring the parameter spaces of binary population synthesis models.

Such statistical studies of XRBs are also useful for the studies of low metal galaxies. In low-metallicity star-forming galaxies, nebular \heii\ emission lines are frequently observed. However, normal stellar populations can not reproduce the observed emission (e.g., \cite{Shirazi2012MNRAS.421.1043S}). This is because the nebular \heii\ emission requires the sources emitting ionizing photons above 54~eV. \citet{Schaerer2019AandA...622L..10S} suggested that X-ray emission from XRBs and ULXs would be the dominant ionizing sources for the nebular \heii\ emission. To understand the contribution of XRBs in \heii\ ionization, quantitative and statistical properties of XRBs in a galaxy and the relation to the host properties such as stellar mass and star formation rate (SFR) are required.

X-ray observatories such as {\it Chandra} \citep{Weisskopf2000}, {\it XMM-Newton} \citep{Jansen2001}, and {\it Swift} \citep{Gehrels2004} collected X-ray photons from numerous celestial objects. Together with those observatories, several studies of XRBs in the extragalactic sky established their X-ray luminosity function (XLF) and the correlation of the integrated X-ray luminosity of XRBs with SFR, stellar mass, and stellar age (e.g., \cite{Grimm2003, Gilfanov2004, Swartz2011, Mineo2012,Zhang2012,Lehmer2014,Peacock2016,Lehmer2019}). Those studies utilized X-ray observations targeting nearby galaxies. 

Today, detected source catalogs of those observatories since their operations are publicly available \citep{Evans2020_CSC2, Webb2020, Evans2020}. The {\it Chandra}, {\it XMM-Newton}, and {\it Swift} catalog contains 317167, 550124, and 206335 unique objects, respectively. {Among these catalogs, {\it Chandra} provides the finest angular resolution down to $\sim0.5$~arcsec.} In this paper, by cross-matching {{\it Chandra} detected} X-ray sources with nearby galaxy catalogs such as the local volume galaxy (LVG) catalog \citep{Karachentsev2013} and {the catalog of the Infrared Astronomical Satellite ({\it IRAS}) survey \citep{Neugebauer1984}}, we aim to make a new XRB catalog and revisit the statistical properties of XRBs in the extragalactic galaxies. 

This paper is organized as follows.  {The {\it Chandra}} X-ray source {catalog is} introduced in \S.~\ref{sec:Xcatalog}. We describe host galaxy catalogs and their properties in \S.~\ref{sec:Gcatalog}. The catalog matching procedure and our XRB catalog are presented in \S.~\ref{sec:XRBcatalog}. Statistical properties of XRB host galaxies and XRBs are shown in \S.~\ref{sec:XRBhost} and \S.~\ref{sec:XRB}, respectively. To understand the statistical properties of XRBs, we describe our binary population synthesis model for comparison with our data in \S.~\ref{sec:SXBPS}. Discussion {including the {\it Swift} and {\it XMM-Newton} catalogs} and conclusions are given in \S.~\ref{sec:discussion} and \S.~\ref{sec:conclusion}, respectively.  Throughout this paper, we adopt the standard cosmological parameters of $(h, \Omega_M, \Omega_\Lambda) = (0.7, 0.3, 0.7)$.

\section{{Chandra Source Catalog}}
\label{sec:Xcatalog}
In this paper, X-ray source information is extracted from the latest {{\it Chandra} source catalog}. The {\it Chandra} X-ray observatory, launched in 1999 July 23, observes the X-ray sky in the energy range between 0.1--10~keV over a field of view of $\sim60$--$250\ {\rm arcmin}^2$. {\it Chandra} achieves a sub-arcsecond on-axis point spread function \citep{Weisskopf2000,Weisskopf2002}. These instrumental capabilities allow to detect sources with low confusion and good astrometry.  On 2019 October 24, a new version of the {\it Chandra} source catalog (CSC2) has been released \citep{Evans2010,Evans2020_CSC2}\footnote{\url{http://cxc.harvard.edu/csc2/}}, which contains 317,167 unique X-ray sources in the sky {including 1,432,324 per observation data sets}. {\footnote{{Note for the Referee, we moved the XMM and Swift catalog part to the discussion section.}}}

{Non-uniform completeness of data sets hamper the construction of unbiased source catalogs. To avoid such biases as much as possible, we select sources from CSC2 as follows. First, multiple observation data would mimic the source properties due to the variability of sources. CSC2 provides both stacked-observation and per-observation detections. The stacked observations compile all the available different-exposure data sets toward sources then stack them, while per-observations provide the information of sources per single observation. Since XRBs are variable sources, we use the per-observation data sets having the longest exposure toward each source.}

{Second, the sensitivity in a field-of-view is not uniform. Generally speaking, sensitivity decreases toward the edge of the field-of-views.} {The detection efficiency of {\it Chandra} is almost uniform down to the flux threshold of $F_{\rm th}=5\times10^{-7}\ {\rm photons\ cm^{-2}\ s^{-1}}$ in the 0.5--7~keV band inside of off-axis angle of 10~arcmin for a $125$~ks exposure (see Fig.~24 of the first CSC catalog paper; \cite{Evans2010}). Thus, we} {select sources having off-axis angle smaller than 10~arcmin.} { Then, we also adopt flux threshold to select samples. As CSC2 detection threshold is improved by a factor of 2 from the first CSC \footnote{\url{https://cxc.cfa.harvard.edu/csc/about.html}}, we select sources brighter than $F\ge (125\ {\rm ks} / t_{\rm obs})F_{\rm th}/2$. $t_{\rm obs}$ is the exposure time. The energy band and spectral assumption are corrected accordingly.} 

{Third, source confusion would bias results. To avoid such a source crowding effect, we select sources not flagged as confused sources.} {  Since XRBs should be compact by definition, we further} { select only compact sources from the catalog \footnote{A few ULXs are also known to be associated with spatially extended X-ray nebulae (e.g., \cite{Cseh2012, Belfiore2020}). These classes of objects will be missed in our catalog.}.}

{ Lastly, the interstellar medium (ISM) absorption will reduce soft X-ray fluxes  (e.g., \cite{Strom1961, Morrison1983, Balucinska1992, Wilms2000}). Detailed spectral analysis is required to decompose the local and Galactic absorption effect. However, such treatments require a certain spectral quality. In this work, for simplicity, we correct the Galactic absorption column density only \citep{Willingale2013}. This treatment would introduce uncertainty in evaluating intrinsic flux at $\lesssim1$~keV. Therefore, we restrict our samples to having the detected photons in the hard band (2-7~keV). 
}


\section{Host Galaxies}
\label{sec:Gcatalog}
We cross-match the observed X-ray sources with two nearby galaxy catalogs: the LVG catalog \citep{Karachentsev2013}\footnote{\url{http://www.sao.ru/lv/lvgdb}} and a catalog of galaxies detected in the {\it IRAS} survey \citep{Neugebauer1984}.

\subsection{LVG Catalog}
\citet{Karachentsev2013} compiled 1075~galaxies in the local volume as the LVG catalog\footnote{In the originally published paper, 869 galaxies were listed.}. The catalog includes galaxies having radial velocities with respect to the centroid of the Local Group $V_{\rm LG}<600\ {\rm km\ s^{-1}}$ or galaxies within the distance of $D_{\rm host}<11$~Mpc. The LVG catalog contains various galaxy characteristics such as angular diameters, axial ratio, apparent magnitudes in various bands, morphological types, and distances. The definition of morphological types of galaxies in LVG is according to \citet{deVaucouleurs1991}. The catalog also provides an estimate of galaxy parameters such as absolute $B$ magnitude and stellar mass estimated via K-band luminosity. We further add infrared flux values at 60~$\mu$m and 100~$\mu$m measured by {\it IRAS} utilizing the NASA/IPAC Extragalactic Database (NED)\footnote{\url{https://ned.ipac.caltech.edu/}} to estimate the SFR based on infrared (IR) measurements as done in \citet{Swartz2011}. 131~LVG galaxies have measurements both at 60~$\mu$m and 100~$\mu$m.

Since we would like to select X-ray sources within a galaxy, the angular diameter, the axial ratio, and the position angle (PA) of a galaxy are needed. The LVG catalog provides the major angular diameter $a_{26}$ defined by the Holmberg isophote $\sim26.5~{\rm mag}\ {\rm arcsec}^{-2}$ in $B$-band and the corresponding apparent axial ratio $b/a$. Some LVG galaxies lack the information on the axial ratio. For those objects, we assume $b/a=1$. PAs are not listed in the original LVG catalog. Therefore, we obtain the PA information from the NED system. The PA information listed in NED is based on the measurement by SDSS-DR6, 2MASS, RC3, and ESO-LV catalogs. When multiple information is available, we adopt the latest available values for each object. As a result, we add PA for 413 LVG galaxies.

\subsection{IRAS Catalog}
The LVG catalog is a complete galaxy catalog, but up to 11~Mpc. To expand the horizon of our catalog, we retrieve IRAS galaxies from the NED database, including information of distance, flux measurements, and size.
{The IRAS sources in the database are based on IRAS Faint Source Catalog, version 2.0 \citep{IRASFSC} and IRAS Point Source Catalog, version 2.0 \citep{IRASPSC}. We select sources with 60 and 100~$\mu$m flux measurements available in either catalog.}
{The distances of the IRAS sources in the database are taken from various literatures. The distance indicator of $\sim70\%$ of the IRAS sources with XRB detection used for the analysis in this paper is based on Tully-Fisher relation. The uncertainty of the corresponding redshift is less than 1\%.}
We restrict galaxies within $z\le0.03\sim130$~Mpc; otherwise, X-ray sources would be biased to only luminous XRBs for distant galaxies. The total number of selected {\it IRAS} galaxies is 11357. We match these {\it IRAS} galaxies with the RC3 catalog \citep{deVaucouleurs1991}, which provides the information of galaxy morphology. The closest objects are selected within the matching radius of 15~arcsec. We note that some of {\it IRAS} galaxies are overlapped with the LVG catalog. In that case, we primarily use the information from the LVG catalog. 

\begin{figure}
 \begin{center}
  \includegraphics[ width=\linewidth]{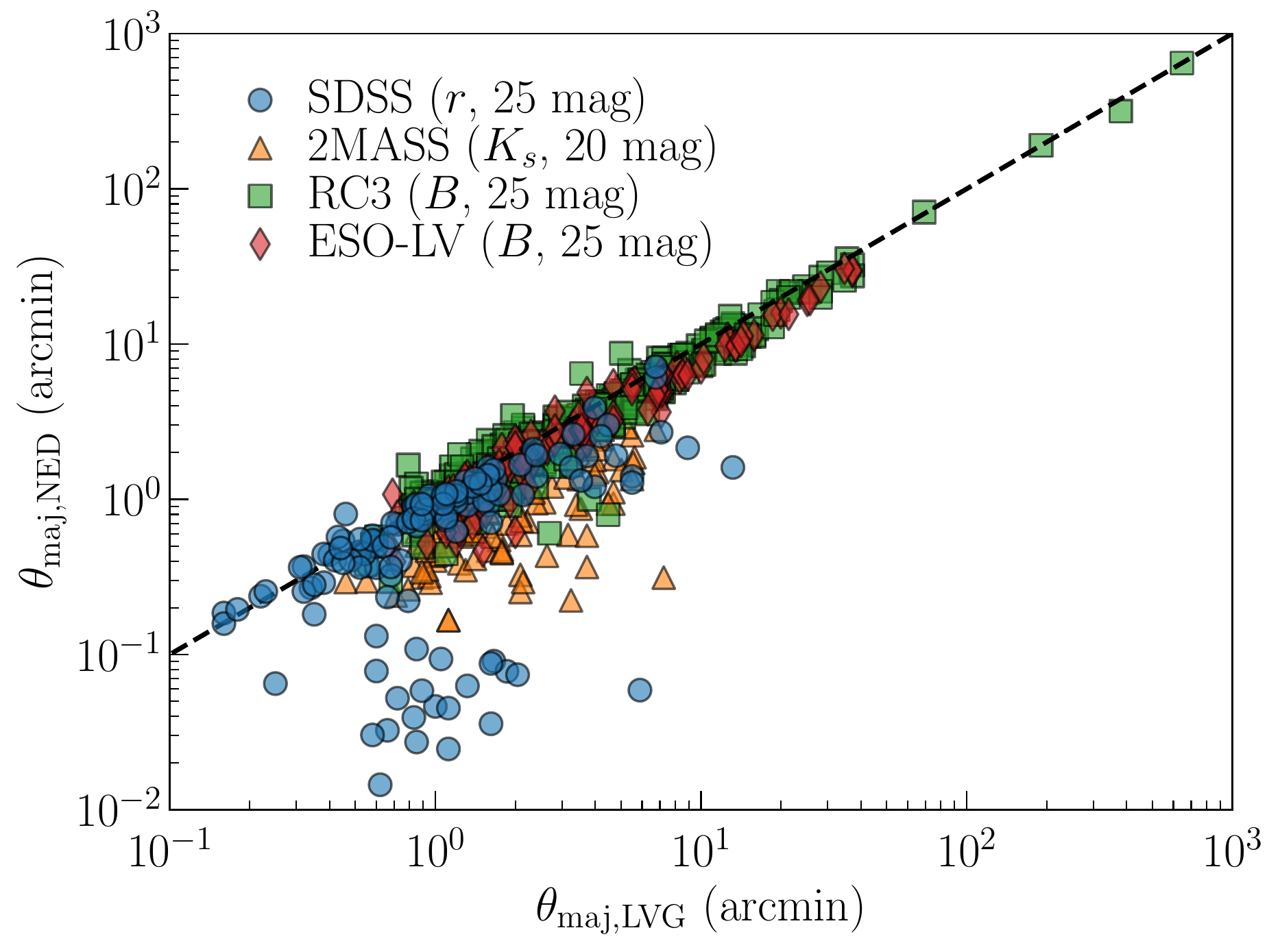}
 \end{center}
\caption{Ratio between major angular diameter of LVG galaxies $a_{26}$ and that reported in the SDSS-DR6, 2MASS, RC3, and ESO-LV catalogs.}\label{fig:a26_lvg}
\end{figure}

The major angular diameter, axial ratio, and PA of {\it IRAS} galaxies are taken from NED, which are based on the SDSS-DR6, 2MASS, RC3, and ESO-LV catalogs. When multiple information is available, we adopt the latest available values. The major diameters for SDSS-DR6, 2MASS, RC3, and ESO-LV catalogs are defined at $r=25$, $K_s=20$, $B=25$, and $B=25$~${\rm {mag\ arcsec}}^{-2}$, respectively. {In literature, RC3 $D_{25}$, defined by the Holmberg isophote $\sim25.0~{\rm mag}\ {\rm arcsec}^{-2}$ in $B$-band, is frequently used. In this paper, we also use $D_{25}$. However, available size information depends on galaxies as described above. Fig.~\ref{fig:a26_lvg} shows the ratio between $a_{26}$ and various other angular estimation. The median ratio between $a_{26}$ and each angular diameter measurement for LVG galaxies is 1.24 (SDSS-DR6), 2.31 (2MASS), 1.07 (RC3), and 1.25 (ESO-LV), respectively.} We take these ratios as conversion factors to correct the values to {$D_{25}$} for the {LVG} and {\it IRAS} galaxies.

\subsection{Properties of External Galaxies}
We estimate the SFR and stellar mass of individual galaxies to compare the number of XRBs against SFRs and stellar masses. In this paper, we adopt the Salpeter initial mass function (IMF; \cite{Salpeter1955}).

\subsubsection{Star formation Rates}
The original LVG catalog provides an estimate of SFR $\dot{\rho}_{\star}$ in a galaxy using H$\alpha$ and the Galaxy Evolution Explorer (\textit{GALEX}) FUV measurements. Following \citet{Kennicutt1998}, H$\alpha$ SFR is estimated as
\begin{equation}
\dot{\rho}_{\star, {{\rm H}\alpha}} = 0.945\times10^9 F^c_{{\rm H}\alpha} d_{\rm host}^2\ [M_\odot\ {\rm yr}^{-1}],
\end{equation}
where $d_{\rm host}$ is the distance to the galaxy in Mpc and $F^c_{{\rm H}\alpha}$ is the extinction corrected integral H$\alpha$ line flux in ${\rm erg\ cm^{-2}\ s^{-1}}$. The H$\alpha$ extinction is $A({\rm H}\alpha)=0.538(A_B^G+A_B^i)$, where $A_B^G$ is the $B$-band extinction according to \citet{Schlegel1998} and $A_B^i$ is the intrinsic $B$-band extinction of the galaxy according to \citet{Verheijen2001}.

Another SFR estimate is based on the measurement of FUV flux of a galaxy. Following \citet{Lee2011}, \citet{Karachentsev2013} used the relation 
\begin{equation}
\dot{\rho}_{\star, \rm FUV} = 2.78-0.4m^c_{\rm FUV}+2\log d_{\rm host},
\label{eq:SFR_FUV}
\end{equation}
where $m^c_{\rm FUV}$ is the extinction corrected FUV magnitude as $m^c_{\rm FUV}=m_{\rm FUV}-1.93(A_B^G+A_B^i)$.

As we combine {\it IRAS} flux measurements, we also estimate the SFR of a galaxy using IR flux measurements following \citet{Kennicutt1998,Swartz2011} as 
\begin{equation}
\dot{\rho}_{\star, \rm IR} = 6.79\times10^{-5} d_{\rm host}^2 (2.58S_{60}+S_{100}),
\end{equation}
where $S_{60}$ and $S_{100}$ is the {\it IRAS} 60 and 100~$\mu$m flux measurement in Jy. However, we note that this estimate reflects only the reprocessed emission component by dust.

\begin{figure*}
 \begin{center}
  \includegraphics[width=0.48\linewidth]{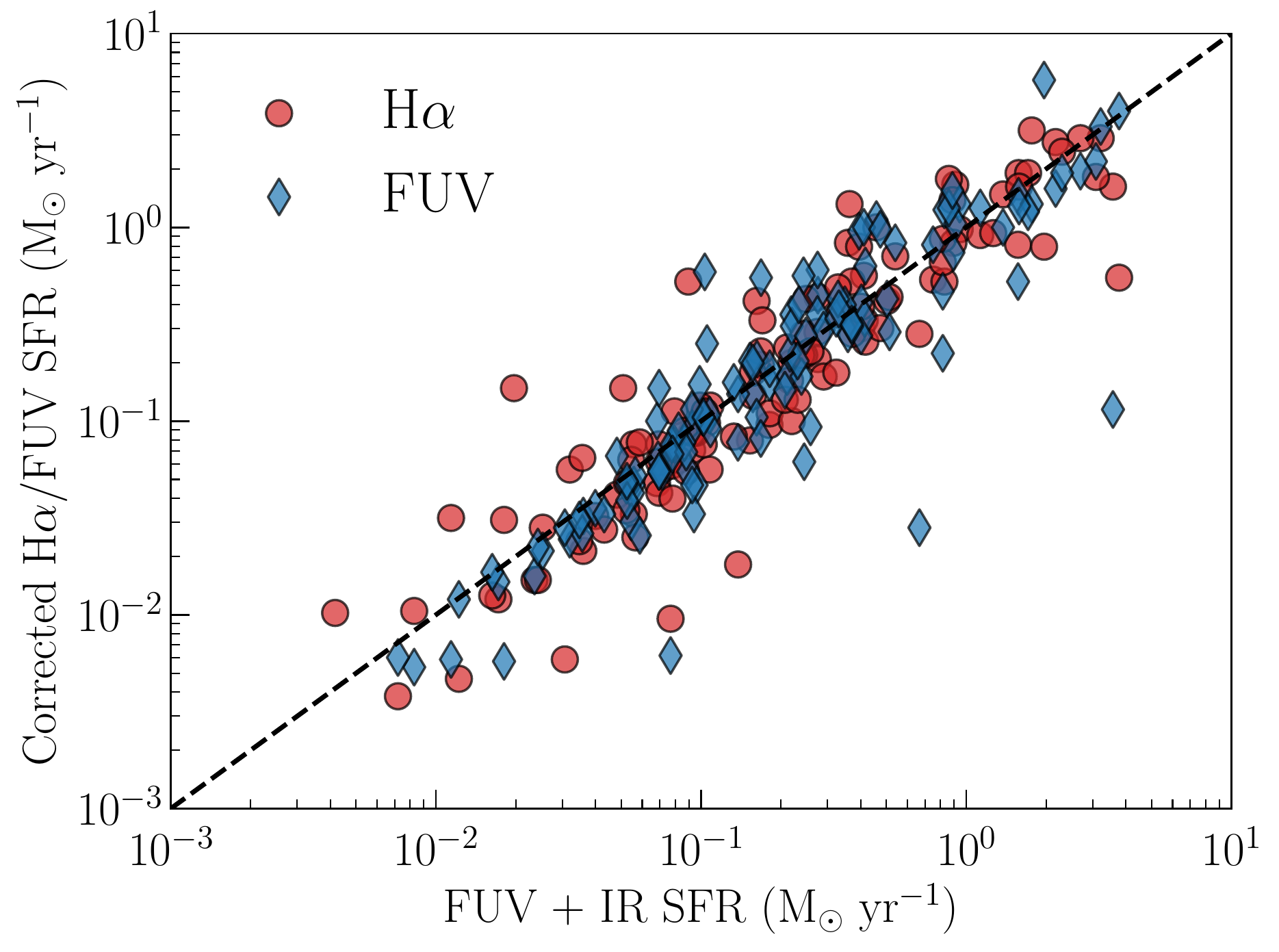}
  \includegraphics[width=0.48\linewidth]{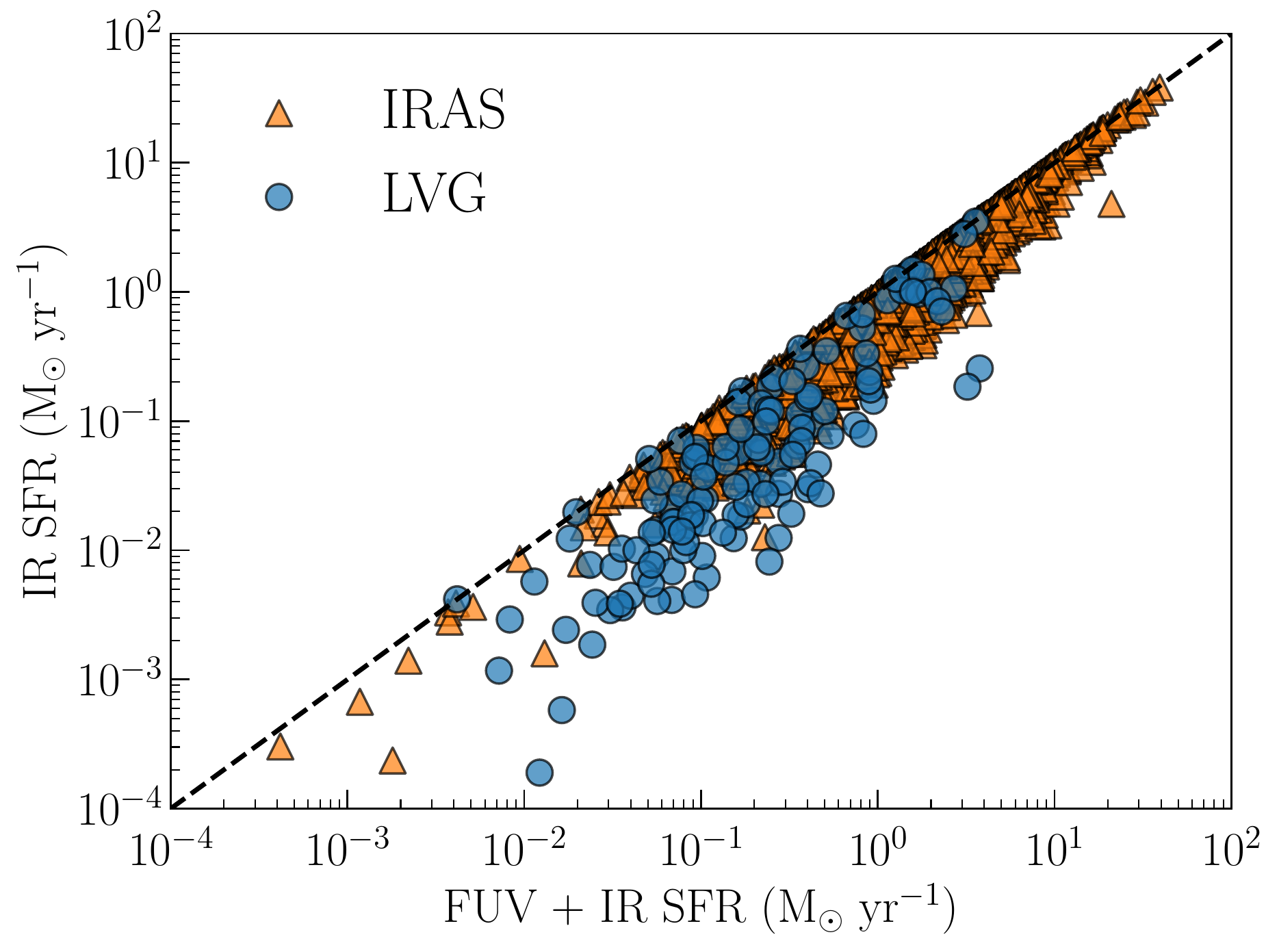}
 \end{center}
\caption{{\it Left}: Comparison of SFR estimates of LVG galaxies between H$\alpha$ and FUV+IR methods (circle). Comparison between FUV and FUV+IR methods are shown in diamond. {\it Right}: Same as the {\it Left} panel, but comparison between IR and FUV+IR methods for LVG galaxies (circle) and {\it IRAS} galaxies (triangle).}\label{figA:SFR}
\end{figure*}

Lastly, in order to avoid the uncertainty of extinction correction, we also make an estimate of the SFR of a galaxy using FUV and IR measurements as
\begin{equation}
\dot{\rho}_{\star, \rm tot} = \dot{\rho}^{e}_{\star, \rm FUV}+\dot{\rho}_{\star, \rm IR},
\end{equation}
where $\dot{\rho}^{e}_{\star, \rm FUV}$ is the FUV SFR estimate but using the FUV flux measurement before correcting internal attenuation.

For {\it IRAS} galaxies, we retrieve {\it GALEX} FUV flux measurement through NED and corrected for the Galactic extinction following \citet{Schlegel1998} as in the LVG catalog where we utilzied the {\tt dustmap} software package \citep{Green2018}. We then derive $\dot{\rho}_{\star, \rm tot}$ of the {\it IRAS} galaxies in the same manner as for the LVG galaxies.

In this paper, we use $\dot{\rho}_{\star, \rm tot}$ as the SFR values ($\dot{\rho}_{\star}$) of both LVG and {\it IRAS} galaxies. If FUV flux information is not available, we adopt $\dot{\rho}_{\star, \rm IR}$. We do not use H$\alpha$ and FUV SFR estimations as H$\alpha$ flux and intrinsic attenuation factor are not available for {\it IRAS} galaxies. Left panel of Fig.~\ref{figA:SFR} shows the comparison of $\dot{\rho}_{\star, \rm tot}$ with $\dot{\rho}_{\star, {{\rm H}\alpha}}$ and $\dot{\rho}_{\star, \rm FUV}$ in 129 LVG galaxies{ , which have both information}. As it can be seen, SFR indicators of H$\alpha$ and FUV may underestimate the total SFR in some galaxies. Right panel of Fig.~\ref{figA:SFR} shows the comparison of $\dot{\rho}_{\star, \rm tot}$ with $\dot{\rho}_{\star, \rm IR}$ in 129 LVG galaxies and 4501 {\it IRAS} galaxies. SFRs of LVG galaxies are more FUV dominated than those of {\it IRAS} galaxies. Even though {\it IRAS} galaxies are IR selected, $\dot{\rho}_{\star, \rm FUV}$ may still contribute to the total SFR at some level. {However, that effect would be around 30\% in $\dot{\rho}_{\star, \rm tot}$, since the median ratio of $\dot{\rho}_{\star, \rm tot}/\dot{\rho}_{\star, \rm IR}$ is 4.2 and 1.3 for LVG and IRAS galaxies, respectively.} 

\subsubsection{Stellar Mass}
For the estimation of stellar mass $M_\star$ of a galaxy, we follow the mass-to-light ($M_\star/L$) relation proposed by \citet{Bell2001}. We adopt their relation for $B$-band and $K$-band measurements. For simplicity, we assume that $K$ and $K_s$ band fluxes are the same. The relation is given as 
\begin{equation}
\log(M_\star /L) = \frac{a_{k} + b_{k}(B-K)}{c_k},
\end{equation}
where $a_k=-0.926$, $b_k=0.205$, and $c_k=0.966$. $B$ and $K$ are absolute magnitudes in the $B$ and $K$ bands. We adopt the formation epoch model with bursts and combine the relations for $B-V$ and $V-K$ in \citet{Bell2001}. Because the stellar mass estimation by \citet{Bell2001} is based on a scaled Salpeter IMF, which is 30\% smaller than that based on the Salpeter IMF, we multiply $1.43$ to the derived stellar mass for our purpose.


\section{Selection of X-ray sources in external galaxies}
\label{sec:XRBcatalog}


\begin{table*}[th]
  \tbl{Description of the Catalog}{%
  \begin{tabular}{ccc}
      \hline
      Column Name & Format & Notes  \\ 
      \hline
Name\_XRB&&X-ray binary name\\
RA & hms & Right ascension of the source\\
DEC & dms & Declination of the source\\
L\_X\_2\_10 & erg~s$^{-1}$ & 2-10 keV luminosity\\
Satellite &  & Identification of the satellite which measured the source$\dag$.\\
Name\_Host &  & Host galaxy name\\
RA\_Host & hms & Right ascension of the host galaxy\\
DEC\_Host & dms & Declination of the host galaxy\\
D$_{25}$ & arcmin & Major angular diameter of the host galaxy\\
b/a &  & Axial ratio between major and minor axes of the host galaxy\\
PA & deg & Position angle of the host galaxy\\
Morphology &  & Morphology of the host galaxy, defined by \citet{deVaucouleurs1991}\\
Distance & Mpc & Distance to the host galaxy\\
SFR & $M_\odot\ {\rm yr}^{-1}$ & Star formation rate of the host galaxy\\
Stellar\_Mass & $M_\odot$ & Stellar mass of the host galaxy\\
Galaxy\_Catalog & & Identification of the parent galaxy catalog$\ddag$\\
      \hline
    \end{tabular}}\label{tab:catalog}
\begin{tabnote}
\footnotemark[$\dag$] C: Chandra, X: XMM-Newton, S: Swift/XRT. 
\footnotemark[$\ddag$] LVG: LVG catalog, IRAS: IRAS catalog. 
\end{tabnote}
\end{table*}



{We} select X-ray sources located in the region of the LVG and {\it IRAS} galaxies where the galaxy size is defined by {$D_{25}$}, $b/a$, and PA. If PA is not available, we define the galaxy region by the circle with a diameter of the square root mean of the major and minor diameters ({$D_{25}$} and $b/a\times {D_{25}}$). We remove largely extended nearby galaxies from our search, LMC ($a_{26}=646^\prime$), SMC ($380^\prime$), M31 ($191^\prime$), {and M33 ($64.7^\prime$)}. We also remove NGC~5195 from the LVG catalog, a merging galaxy with NGC~5194, also known as M~51, and {$D_{25}$} of NGC~5194 includes the counterpart galaxy NGC~5195. 

By using the distance information of host galaxies, we calculate the intrinsic X-ray flux and luminosity in the 2-10~keV band of XRB candidates. Intrinsic fluxes and luminosities are estimated as follows. {The CSC2 provides the observed flux in the 2.0-7.0~keV  of each object.}  We convert those fluxes to the intrinsic luminosities by assuming an absorbed power-law spectrum with photon index $\Gamma=1.8$, which is the typical photon index of XRBs \citep{Swartz2004, Swartz2011}. Even if we set it as $\Gamma=1.7$ or $1.9$, the XLF results do not change significantly. Distances to the objects set as the distance to their associated host galaxies. The absorption due to the Galactic cold interstellar medium is modeled using the {\tt TBabs} code \citep{Wilms2000}, in which cross sections of dust grains and molecules are taken into account. The absorption hydrogen column density ($N_{\rm H}$) of each line-of-sight is fixed to the value estimated by \citet{Willingale2013}, in which the contribution of not only neutral hydrogen atoms ($N_{\rm HI}$) but also molecular hydrogen ($N_{\rm H2}$) are included.


{The} X-ray sky is known to be dominated by active galactic nuclei (AGNs; see e.g., \cite{Ueda2014}). We expect significant contamination of background AGNs in the X-ray source catalogs. To remove known AGNs from the catalog, we cross-match the X-ray catalogs with available AGN catalogs covering a wide area of the sky. The {\it Wide-field Infrared Survey Explorer} (WISE; \cite{Wright2010}), observing the {\it IR} sky, is one of the ideal missions to identify a huge number of AGNs across the full sky. The {\it WISE} AGN catalog contains 4,543,530 AGN candidates with 90\% reliability \citep{Assef2018}. AGN candidates are selected by the {\it IR} color information. \citet{Veron2010} also provides a catalog of AGNs containing 168,941 AGNs in its 13th Edition. We utilize these two catalogs to reject the contamination of known AGNs. Following the same methods for the X-ray catalog comparison above, we collect all the X-ray sources around each AGN catalog within 60~arcsec. Based on the distributions, we define the tolerance radii, where the chance coincidence becomes 5\%. We remove those matched X-ray sources, which can make about {7\% of the number of X-ray sources. In addition, central objects can be contamination of low-luminosity AGNs or unresolved sources. Therefore, we remove sources whose positional error is within galactic center.}

The surface number density of the faintest X-ray sources reaches $\approx 50500\ {\rm deg^{-2}}$, dominated by AGNs, based on the {\it Chandra} deep field survey with $\sim7$~Ms observations \citep{Luo2017ApJS..228....2L}, which goes down to  $4.2\times10^{-18}$ and $2.0\times10^{-17}\ {\rm erg\ cm^{-2} s^{-1}}$ at 0.5-2 and 2-7~keV, respectively. The median size of the angular diameter of the selected LVG and the {\it IRAS} galaxies are {$8.1^\prime$} and {$4.2^\prime$}, respectively. Thus, there may be contamination of background unknown AGNs even after removing known AGNs. To avoid contamination of such unknown AGNs to the XRB catalog, we remove galaxies whose X-ray source number density is below {the} AGN surface density fluctuation in all the flux regions. {We note that the} 1-$\sigma$ flux fluctuation of the cosmic X-ray background radiation in the 2-10~keV band is $6.49^{+0.56}_{-0.61}$\% \citep{Kushino2002}. We adopt the AGN 2-10~keV source count distribution of \citet{Ueda2014}. Distant galaxies and high-latitude stars would also contaminate. At $>2$~keV, however, those populations are insignificant in the source count comparing to AGNs \citep{Lehmer2012}.

The sensitivity of current X-ray observatories allows us to see even extragalactic supernova remnants (SNRs). SNRs may start to become a main X-ray source population at $\lesssim10^{35.5}\ {\rm erg\ s^{-1}}$ at 0.5-2~keV (e.g., \cite{Binder2012}).  Although SNRs tend to have soft spectra, we restrict our samples whose X-ray luminosity is $\ge10^{36}\ {\rm erg\ s^{-1}}$ to pick up XRBs. We note that some SNRs are known to be as bright as $10^{37}\ {\rm erg\ s^{-1}}$ in X-rays \citep{Ghavamian2005,Maggi2016}. {Since the highest ULX luminosity is $\sim10^{42}\ {\rm erg\ s^{-1}}$ \citep{Farrell2009Natur.460...73F}, we restrict samples whose X-ray luminosities do not exceed $10^{42}\ {\rm erg\ s^{-1}}$.}

\begin{figure}[t]
 \begin{center}
  \includegraphics[width=\linewidth]{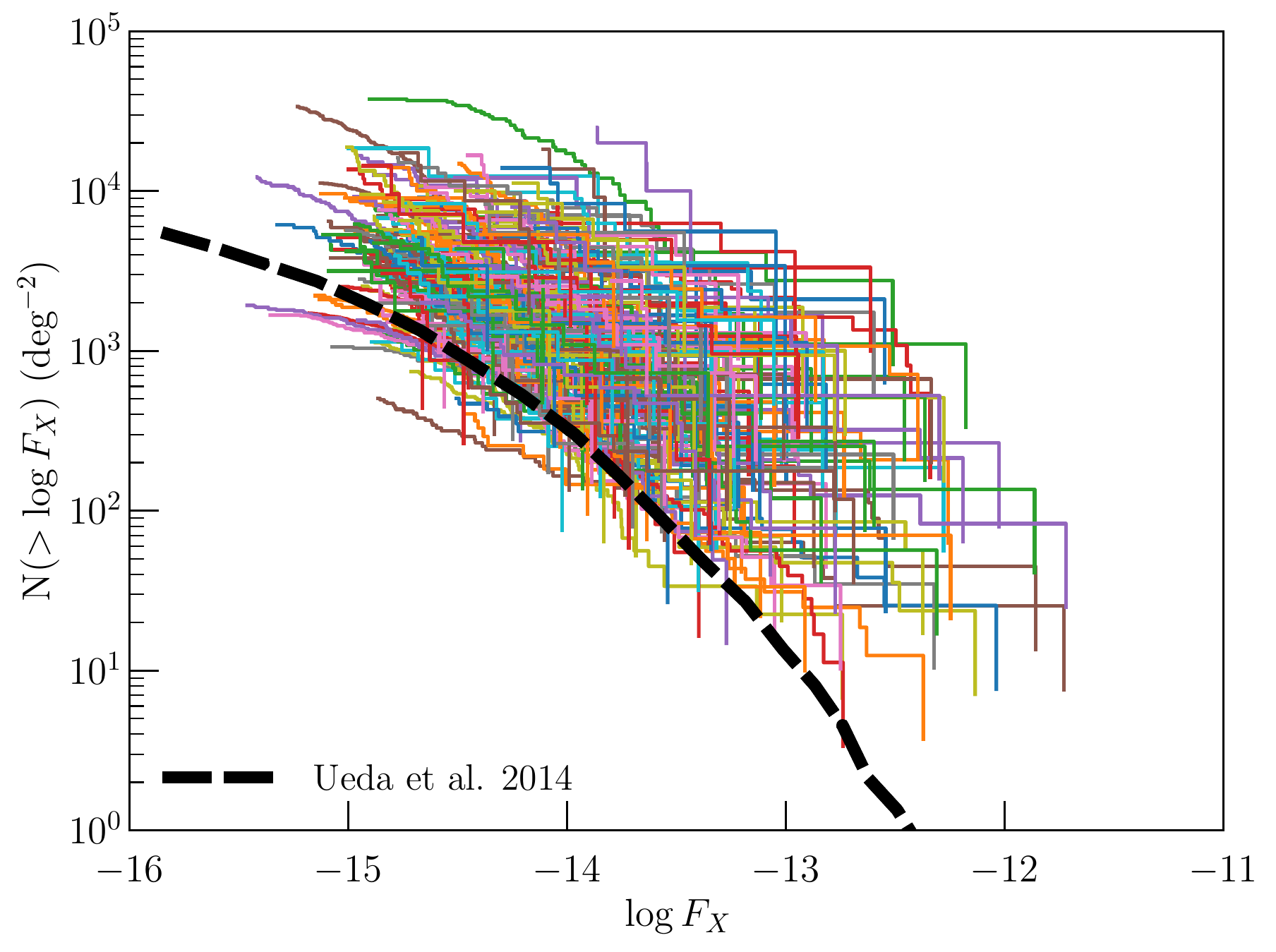}
 \end{center}
\caption{Cumulative source count distribution of XRBs in the LVG and {\it IRAS} galaxies in the 2-10~keV band. The source count is normalized by the area size of the galaxies. Each solid line corresponds to each LVG galaxy. The dashed line represents the source count distribution of background AGNs taken from \citet{Ueda2014}.}\label{fig:logN_logS_flux}
\end{figure}

Finally, after applying the criteria above, our catalog includes {4430} X-ray sources associated with {237} galaxies. Among those XRB candidates, {378} are ULX candidates. Hereinafter, we treat those XRB candidates as XRBs. The catalog is available from the following link \url{http://astro-osaka.jp/inoue/page/exrbcatalog/}. The description of the catalog is presented in Table~\ref{tab:catalog}. {As described in \S~8.4, we also include {\it Swift} and {\it XMM-Newton} sources in this catalog. However, in this paper, we rely on {\it Chandra} detected sources, otherwise noted.} We also provide a list of extended objects, although they are not used in our study. Figure. \ref{fig:logN_logS_flux} shows the cumulative source count distribution of XRBs in the selected LVG and {\it IRAS} galaxies in the 2-10~keV band. For comparison, the source count distribution of AGNs is also shown \citep{Ueda2014}.

\begin{figure}
 \begin{center}
  \includegraphics[width=\linewidth]{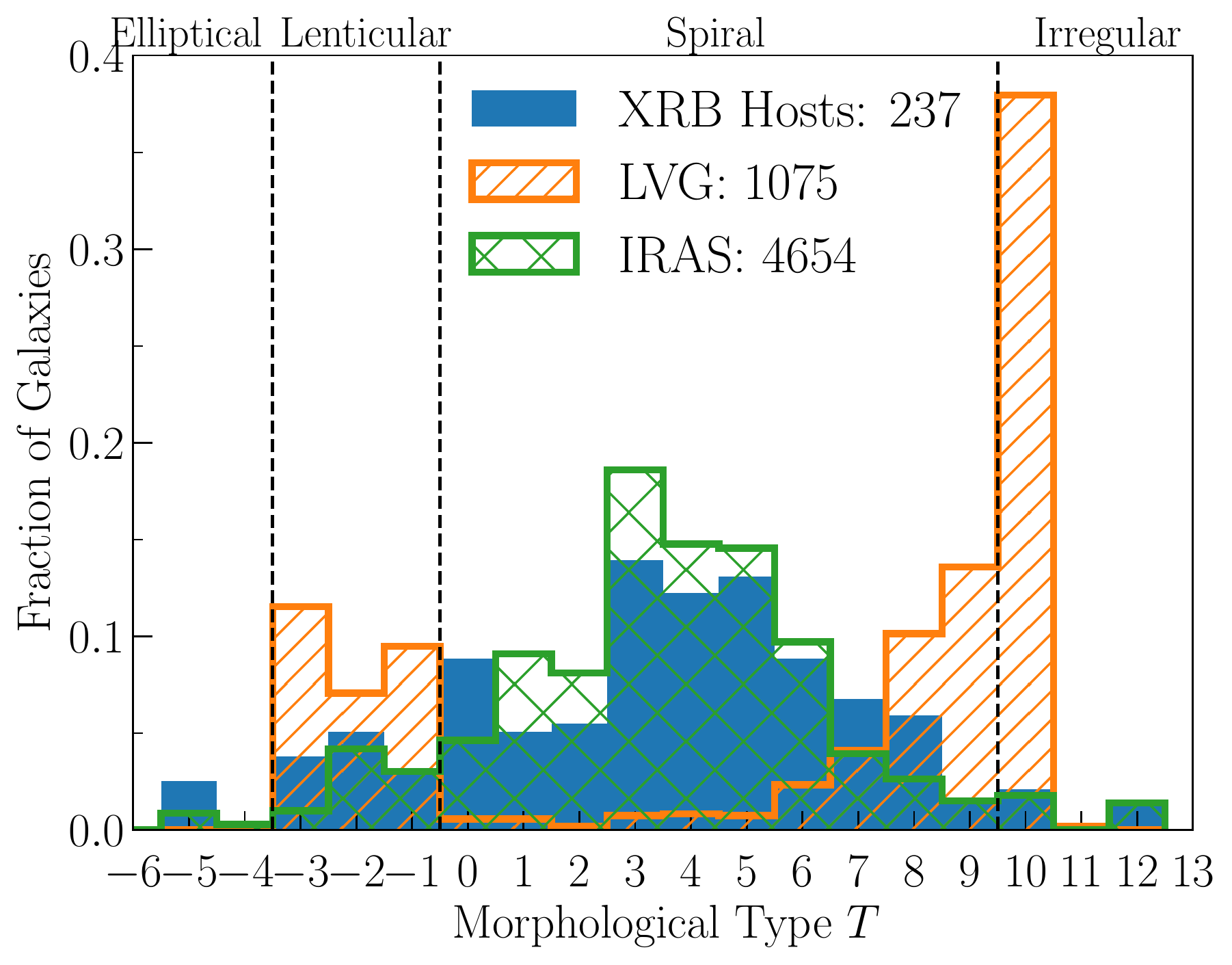}
 \end{center}
\caption{Distribution of morphological types of XRB hosting galaxies. The distributions of parent galaxy samples LVG and IRAS are also shown. The distributions are normalized to unity. The vertical lines correspond to the divisions of galaxy types. The definition of morphological type follows \citet{deVaucouleurs1991}, but we set peculiar galaxies as $T=12$ for illustrating purpose.}\label{fig:host_type}
\end{figure}
\section{Properties of XRB Host Galaxies}
\label{sec:XRBhost}

\begin{figure}
 \begin{center}
  \includegraphics[width=\linewidth]{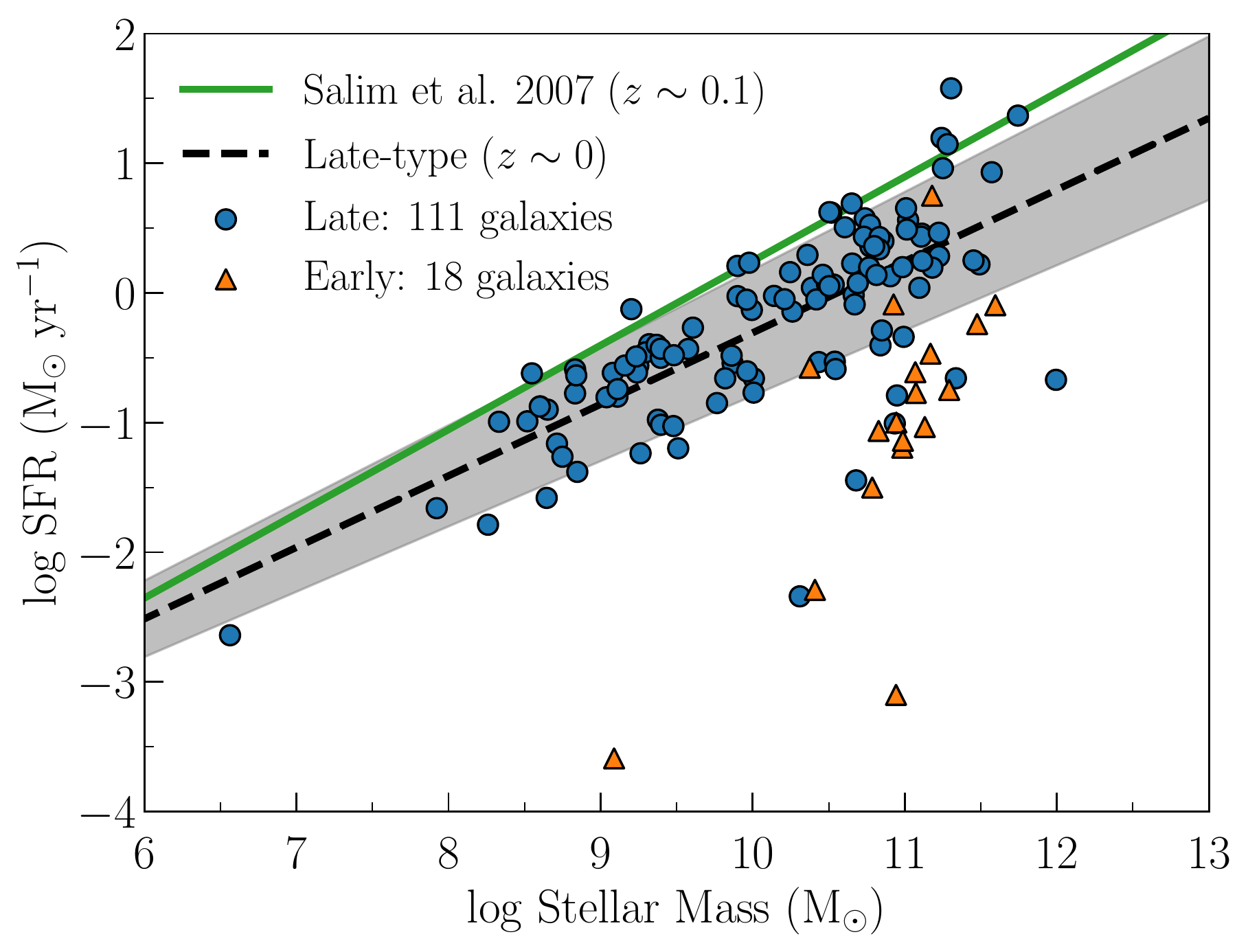}
 \end{center}
\caption{Comparison between stellar mass and SFR of galaxies in XRB hosting  galaxies. Circle and triangle data points correspond to late-type and early-type galaxies, respectively. Dashed line shows the linear regression fit to the late-type galaxies with 1-$\sigma$ error region (shaded region). Solid line is the main-sequence relation at $z\sim0.1$ \citep{Salim2007}.}\label{fig:SM_SFR_XRB} 
\end{figure}

In this section, we discuss the properties of the {237} XRB host galaxies. Fig.~\ref{fig:host_type} shows the normalized distribution of morphological type $T$ of XRB hosting galaxies. We follow the definition of morphological types in Table.~2 of \citet{deVaucouleurs1991}. However, for the plotting purpose, we set that $T=12$ corresponds to peculiar galaxies. As clearly seen in the Figure, late-type galaxies ($T\ge0$) dominate the XRB hosting galaxies in our samples. {$\sim85$\%} of the XRB hosting galaxies are late-type. For comparison, we also show the normalized distribution of $T$ of the whole LVG and IRAS galaxies. Since the LVG catalog includes a number of local dwarf galaxies, their distribution becomes different from the others.

\begin{figure*}
 \begin{center}
  \includegraphics[width=0.48\linewidth]{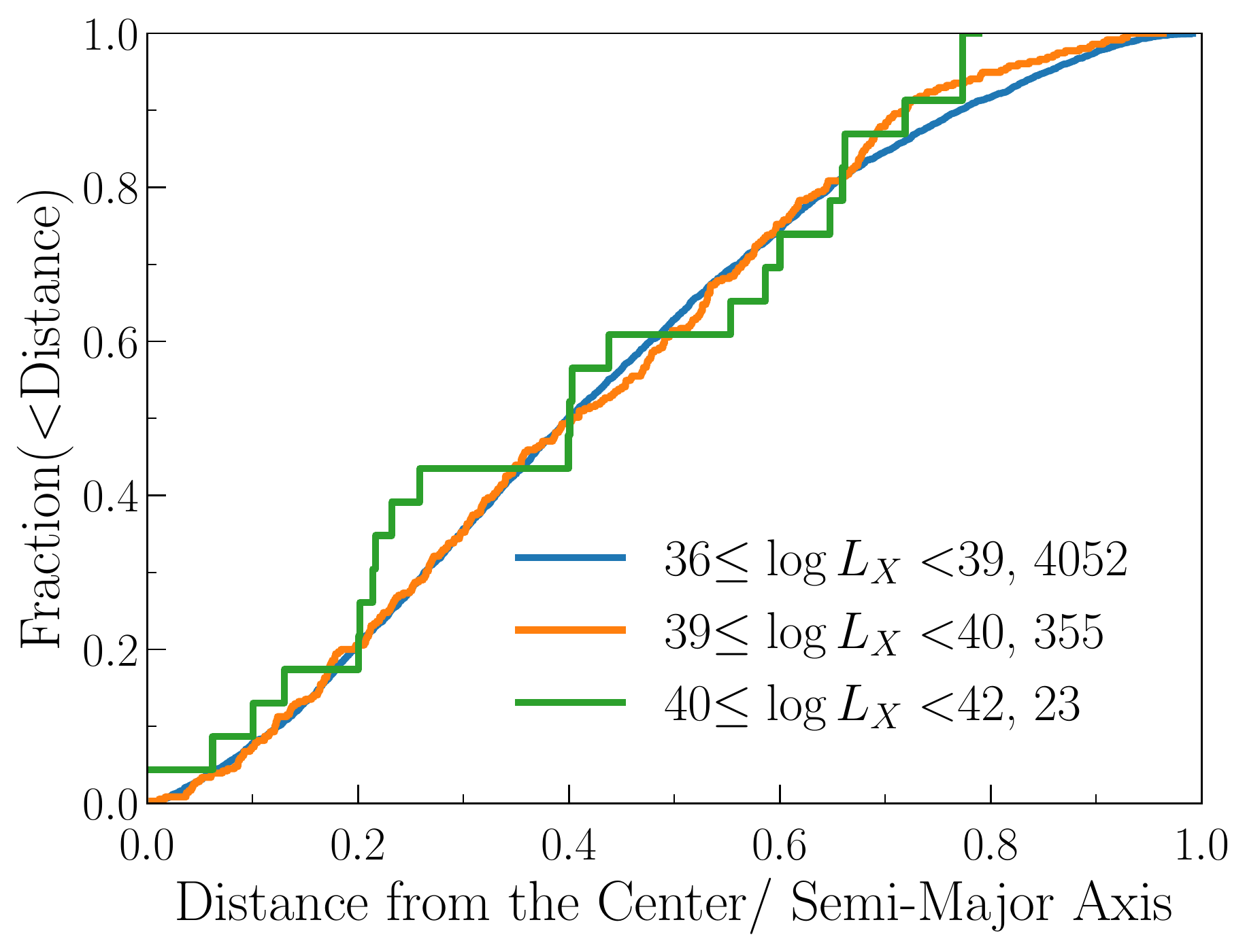}
  \includegraphics[width=0.48\linewidth]{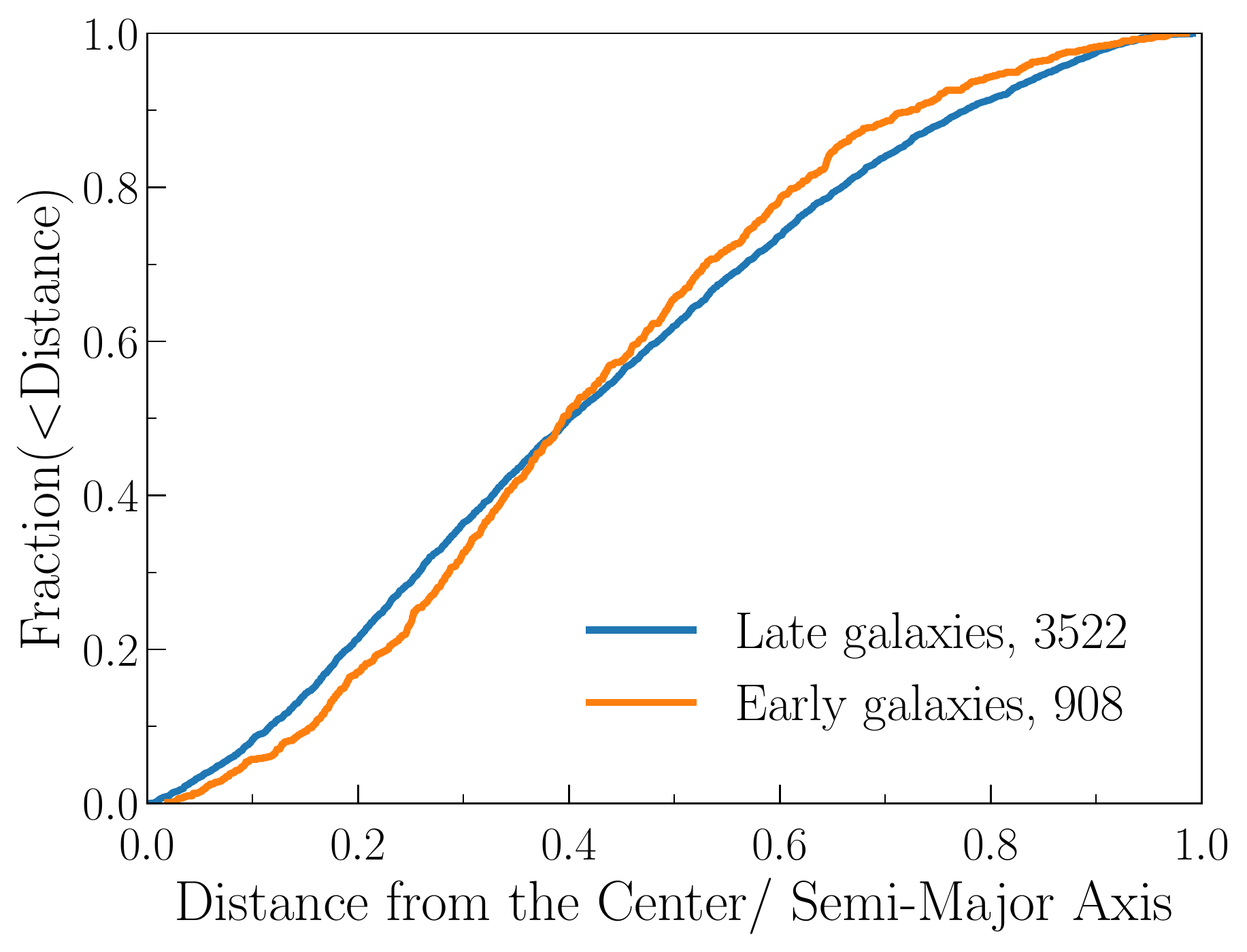}
 \end{center}
\caption{{\it Left}: Radial distribution of XRBs in a cumulative form. The radial distance of each source is normalized by the semi-major axis of their host galaxies. Each curve corresponds to each luminosity bin as presented in the figure. Number of samples included in each curve is also shown in the figure. {\it Right}: Same as the left panel, but showing the distributions depending on galaxy types.}\label{fig:radial_xrb} 
\end{figure*}

We show the relation between stellar mass and SFR of XRB hosting galaxies in Fig.~\ref{fig:SM_SFR_XRB}. Since we restrict host galaxies having both stellar mass and SFR, the number of the total remaining galaxies is reduced to {129} in this Figure. As seen in the plot, heavier XRB host galaxies have higher SFR. This is similar to a well-known galaxy main sequence (e.g., \cite{Noeske2007, Salim2007}), which is defined by a relatively steady star-formation rate in disk galaxies. One would like to quantify the relation between SFR and stellar mass to compare XRB hosting galaxies with other studies. The sequence for the late-type galaxies is best described with a linear fit:
\begin{equation}
	\log \dot{\rho}_{\star} = {(-0.31\pm0.05)} + {(0.55\pm0.05)}\times (\log M_\star - 10),
\end{equation}
which is shown in the Figure. This relation is similar to that of galaxies in the nearby Universe (see, e.g., \cite{Salim2007}). Slight offset comparing to \citet{Salim2007} for $z\sim0.1$, after the IMF correction, may be due to the redshift evolution. When we include early-type galaxies, the correlation changes as 
\begin{equation}
	\log \dot{\rho}_{\star} = {(-0.46\pm0.06)} + {(0.46\pm 0.06)}\times (\log M_\star - 10).
\end{equation}
The slope index is consistent with that for the late-type galaxies within the uncertainty. This may be due to a small number of XRB hosting early-type galaxies.

\section{X-ray Source Distribution}
\label{sec:XRB}
In this section, we discuss the statistical properties of those XRBs.

\subsection{Radial Distribution}

Radial distribution of XRBs gives an independent critical test for our selection as XRBs are off-nucleus objects. The left panel of Fig.~\ref{fig:radial_xrb} shows normalized cumulative radial distributions of XRB candidates in various X-ray luminosity bins of the 2-10~keV X-ray luminosity, $L_X$. About {0.2}\% of the sources are located within the scale of 1\% of the semi-major axis. As no apparent central clustering exists, most of our samples can be regarded as off-nucleus objects. We perform the Kolmogorov-Smirnov (KS) tests to check whether the distributions among luminosities are statistically different. {For all the luminosity bins of $36\le\log L_X<39$, $39\le\log L_X<40$, and $40\le\log L_X<42$, we can not reject the null hypothesis. Further detailed investigations on the spatial distribution of XRBs and ULXs would be required for the understanding of this trend.}

{It has been discussed that HMXBs are the dominant population in late-type galaxies, while LMXBs are in early-type galaxies (see, e.g., \cite{Grimm2003, Gilfanov2004}).} The right panel of Fig.~\ref{fig:radial_xrb} also shows the cumulative radial distributions, but divided into two different galaxy types, late ($T\ge 0$) and early ($T<0$). {The obtained KS-value is $6.3\times10^{-2}$ with a p-value of $0.006$. This comparison implies statistically different spatial distributions between galaxy types, corresponding to the distributions of HMXBs and LMXBs.}

\begin{figure}
 \begin{center}
  \includegraphics[width=\linewidth]{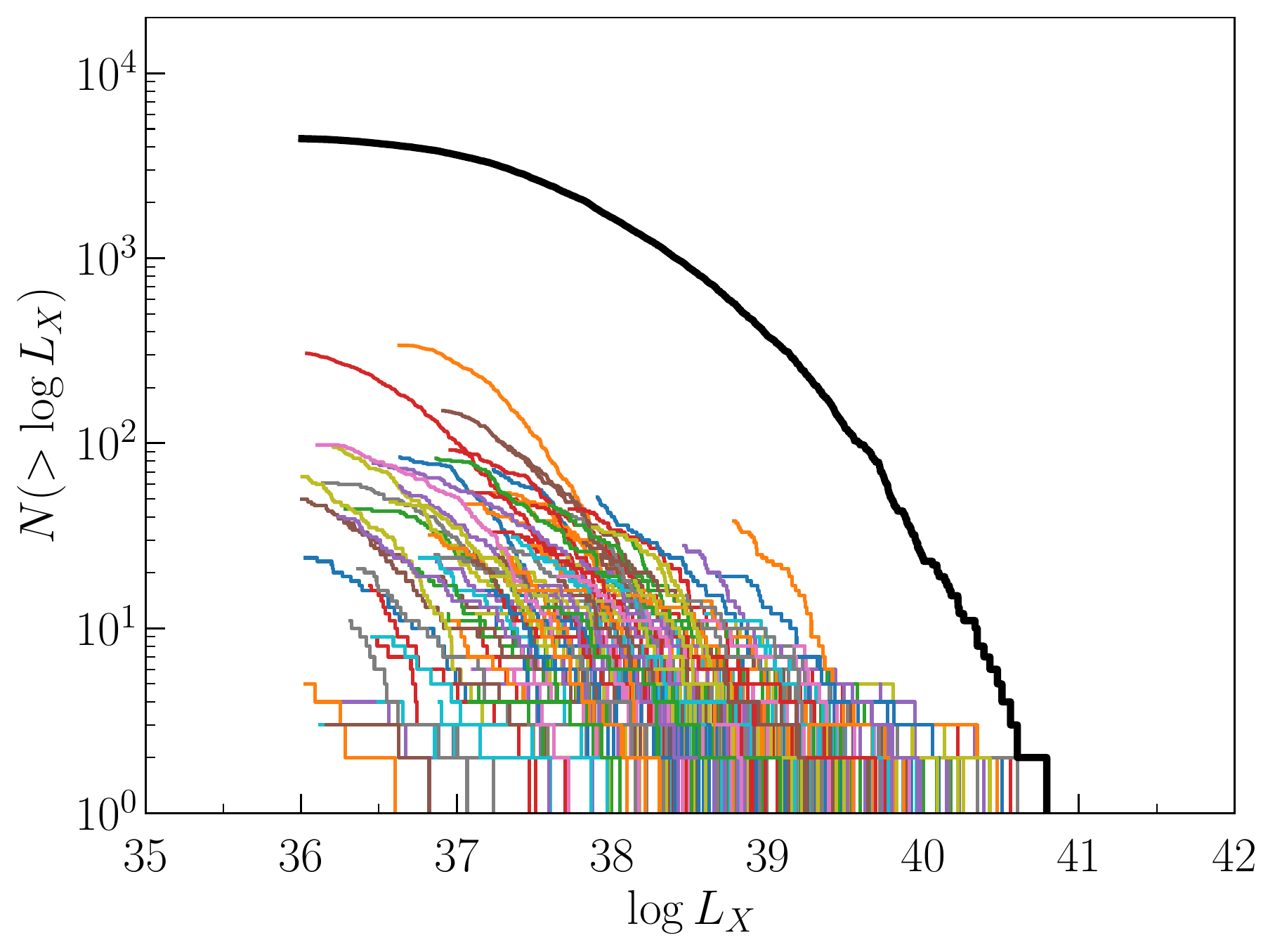}
 \end{center}
\caption{Cumulative source count distribution in luminosity of XRBs of each member galaxy in our samples. The solid line represents the summation of the whole galaxy sample. The luminosity is defined in the 2-10~keV band.}\label{fig:logN_logL} 
\end{figure}

\subsection{Star Formation Rate Normalized X-ray Luminosity Function}
\label{sec:SFRXLF}

\begin{figure}
 \begin{center}
  \includegraphics[width=\linewidth]{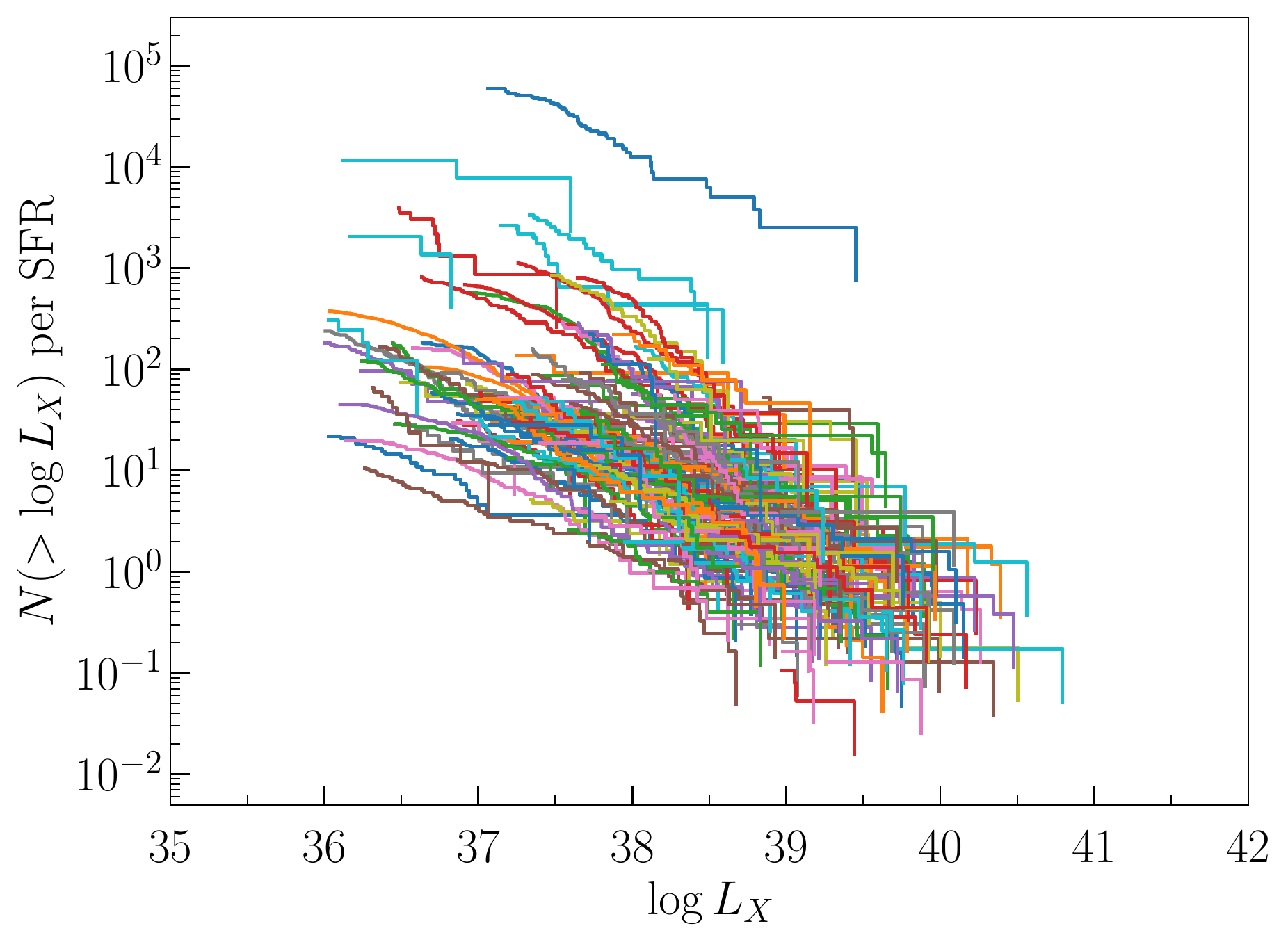}
 \end{center}
\caption{Cumulative luminosity distribution of XRBs of each member galaxy in our samples. The distribution is normalized by star formation rates of each galaxy. The luminosity is defined in the 2-10~keV band.}\label{fig:logN_logL_All} 
\end{figure}

Fig.~\ref{fig:logN_logL} shows the cumulative source count distribution in luminosity of each member galaxy in our samples. We also show the cumulative distribution of all the XRBs in our catalog. The distribution spreads in wide range, and no clear trends seem to exist. Here, the number of XRBs is known to scale with SFR of the host galaxy (e.g., \cite{Grimm2003, Swartz2011, Mineo2012}). Fig.~\ref{fig:logN_logL_All} shows SFR normalized cumulative luminosity count distributions of each member galaxy in our samples where SFR information is available. The overall shape of cumulative luminosity distribution is similar to each other. 

The galaxy having the highest SFR normalized source count is NGC~3379, which is a nearby elliptical galaxy ($T=-3$) at 11~Mpc away with SFR and stellar mass of $8\times10^{-4}\ {\rm M_\odot\ yr^{-1}}$ and $8.8\times10^{10}\ {\rm M_\odot}$, respectively. 

\begin{figure}
 \begin{center}
  \includegraphics[width=\linewidth]{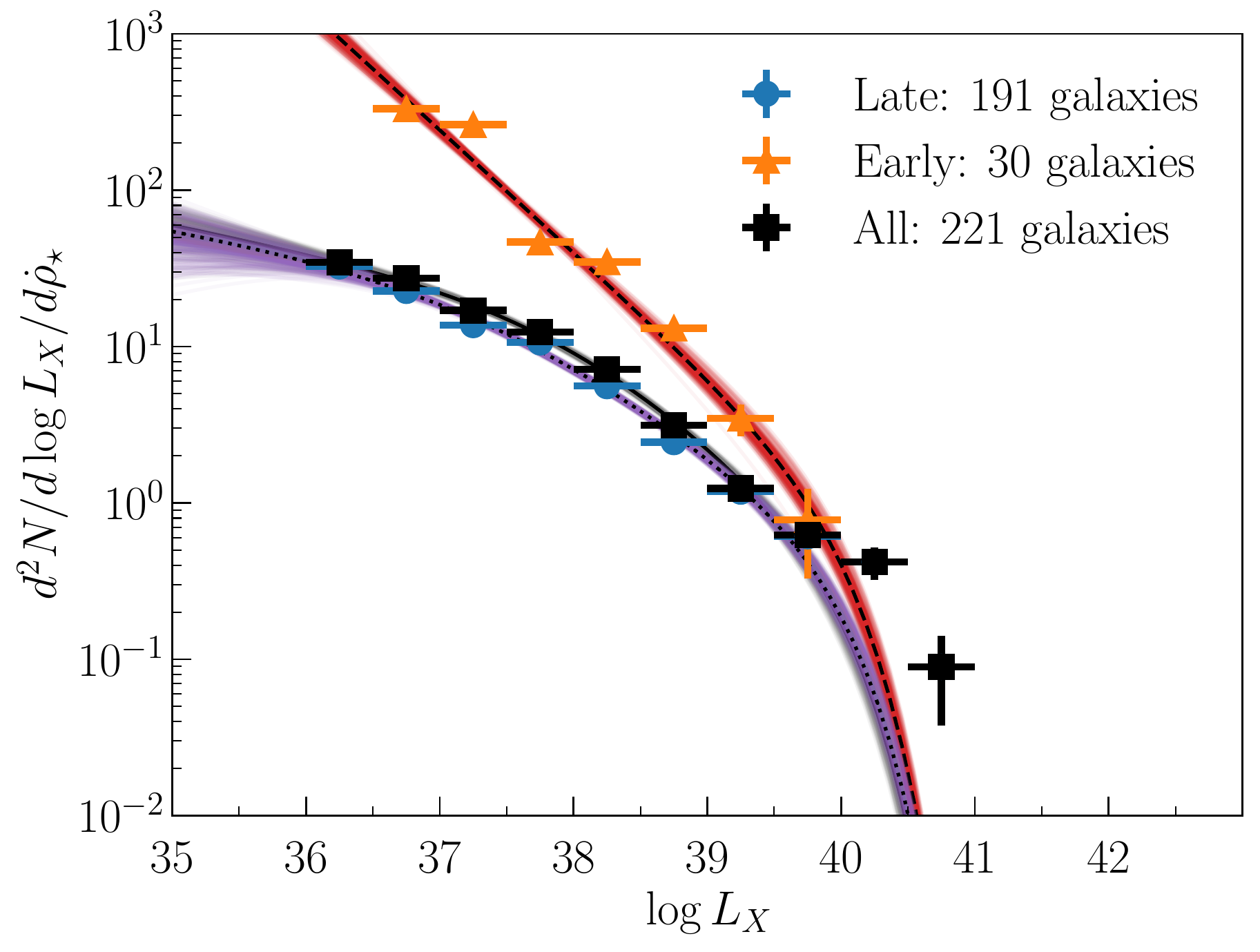}
 \end{center}
\caption{Star formation rate normalized X-ray luminosity function of XRBs in nearby galaxies. The circle, triangle, and square points correspond to that of the late-type, early-type, and all galaxies, respectively. Dotted, dashed, and solid line shows the fitted SFR normalized XLF model for the late-type, early-type, and all galaxies, respectively. The shaded region corresponds to its error region. The models for late-type and all galaxies overlay each other.}\label{fig:dN_dlogL_Type} 
\end{figure}

Fig.~\ref{fig:dN_dlogL_Type} shows the SFR normalized XLF, ${d^2N}/{d\log L_Xd\dot{\rho}_\star}$, obtained by taking the median of the combining data of full galaxy samples. Source luminosity is in the range of $\log L_X\ge 36$. We divide our samples into the late-type galaxies, $T\ge0$, and early-type galaxies, $T<0$. As most of our galaxy samples are late-type galaxies, the distributions of the late-type and the full samples look comparable.

In order to understand statistical properties of SFR normalized XLFs, we fit the binned SFR normalized XLF data. We adopt a model having a broken power-law with a cut-off  for the fit. The XLF model is given in the form of
\begin{equation}
\frac{d^2N}{d\log L_Xd\dot{\rho}_\star} =   A \left[\left(\frac{L_X}{L_b}\right)^{\gamma_1}+\left(\frac{L_X}{L_b}\right)^{\gamma_2}\right]^{-1}\exp{\left(-\frac{L_X}{L_c}\right)}.
\label{eq:XLF}
\end{equation}
{When the broken power-law is not able to fit the binned XLF data, we adopt a single power-law form (i.e., removing the term of $\gamma_1$ in the XLF).}

Since we can not determine the cut-off luminosity $L_c$ from the XLF fit, we fix it as {$\log L_c=40.0$. Dependence on $L_c$ will be discussed in \S.~\ref{sec:FP}}. We perform Markov chain Monte Carlo (MCMC) fitting in order to constrain parameters by using the {\tt emcee} package \citep{Foreman-Mackey2013}. We assume flat distributions for priors of parameters. The fitting results are shown in Table.~\ref{tab:XLF} {together with likelihood values} and Fig.~\ref{fig:dN_dlogL_Type}. For the error region, we use the highest posterior probability density interval containing 68\% of the walker samples.  

{The XLF of late-type galaxies shows flattening in the low luminosity regime. This flattening might be due to the incompleteness of observatories. In this paper, even though we select sources at off-axis angles smaller than 10~arcmin} {and sources whose flux is above the flux threshold having uniform completeness} {, we do not correct the XLF for possible survey incompleteness (i.e., flux dependence of the survey area) of each {\it Chandra} observation.} Later, we discuss a possible cause of this break by comparing it with a phenomenological binary evolution model. We note that, since the total X-ray luminosity is dominated by the high-luminosity end considering the obtained $\gamma_2${, since $\gamma_2 \le 1$. Therefore,} the faint-end slope uncertainty does not affect the estimation of the total luminosity.

\begin{table*}
\tbl{Best-fit parameters for the SFR ($\dot{\rho}_\star$), stellar mass ($M_\star$), and fundamental plane ($\mu$) normalized XLF models.}{
\begin{tabular}{cccccc}
  \hline              
  {Galaxy Type} & {$A$} & {$\gamma_1$} & {$\gamma_2$} & {$\log L_b$} & {log likelihood} \\
  \hline
\multicolumn{6}{c}{SFR ($\dot{\rho}_\star$) Normalized XLF}\\
\hline
All & $22.03_{-7.02}^{+9.56}$ & $0.15_{-0.07}^{+0.06}$ & $0.75_{-0.07}^{+0.08}$ & $37.83_{-0.38}^{+0.35}$ & $-7.177$\\
Late & $34.53_{-14.31}^{+15.64}$ & $0.09_{-0.12}^{+0.09}$ & $0.62_{-0.05}^{+0.07}$ & $37.10_{-0.49}^{+0.59}$ & $-11.469$\\
Early & $242.95_{-12.28}^{+12.14}$ & --- & $0.78_{-0.03}^{+0.03}$ & $37.0$ (fixed) & $-60.499$\\
\hline
\multicolumn{6}{c}{Stellar Mass ($M_\star$) Normalized XLF}\\
\hline
All & $ 6.91_{-0.15}^{+0.15}\times10^{-10}$ & --- & $0.55_{-0.01}^{+0.01}$ & $37.0$ (fixed) & $-78.187$ \\
Late & $ 8.84_{-0.22}^{+0.22}\times10^{-10}$ & --- & $0.56_{-0.01}^{+0.01}$ & $37.0$ (fixed) & $-38.562$\\
Early & $2.24_{-0.09}^{+0.10}\times10^{-10}$ & --- & $0.55_{-0.02}^{+0.02}$ & $37.0$ (fixed) & $6.071$\\
\hline
\multicolumn{6}{c}{Fundamental Plane ($\mu$) Normalized XLF}\\
\hline
All & $13.33_{-4.02}^{+5.04}$ & $0.10_{-0.09}^{+0.07}$ & $0.73_{-0.06}^{+0.07}$ & $37.37_{-0.36}^{+0.35}$ & $ -30.549$\\
Late & $17.58_{-4.90}^{+5.00}$ & $0.01_{-0.12}^{+0.09}$ & $0.65_{-0.06}^{+0.07}$ & $ 37.09_{-0.35}^{+0.38}$ & $-11.698$\\
Early & $4.54_{-0.26}^{+0.26}$ & --- & $0.50_{-0.03}^{+0.03}$ & $37.0$ (fixed) & $-82.074$\\
\hline
\end{tabular}}\label{tab:XLF}
\begin{tabnote}
Errors are in 1-$\sigma$ uncertainty region. {We fix $\log L_c=40.0$.}
\end{tabnote}
\end{table*}

\subsection{Stellar Mass Normalized X-ray Luminosity Function}
\label{sec:SMXLF}

\begin{figure}
 \begin{center}
  \includegraphics[width=\linewidth]{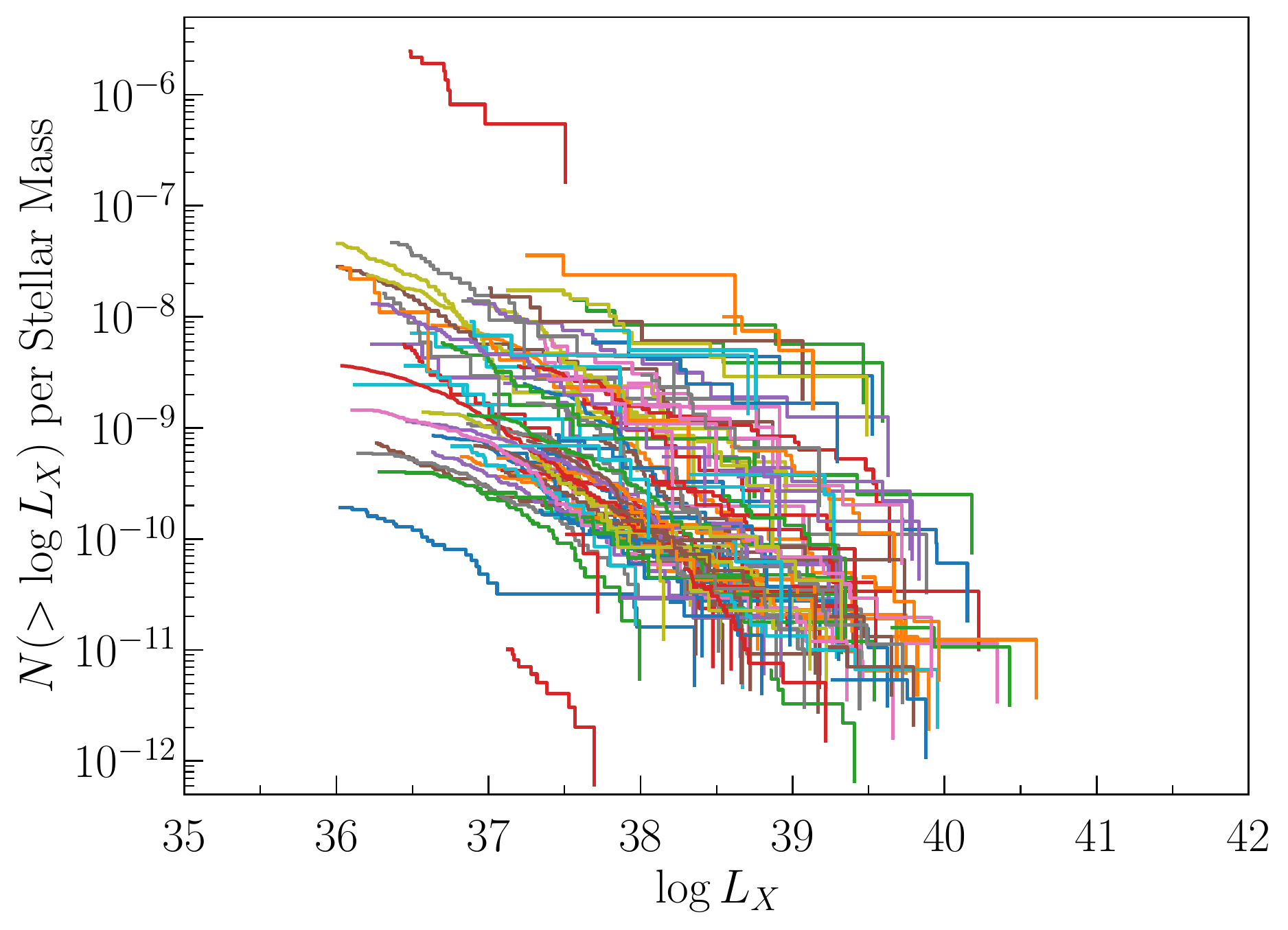}
 \end{center}
\caption{Same as in Fig.~\ref{fig:logN_logL_All}, but normalized by stellar mass of each galaxy.}\label{fig:logN_logL_SM_All} 
\end{figure}

The number of XRBs is also known to scale with their stellar mass of the host galaxy, especially in early-type galaxies (e.g., \cite{Gilfanov2004, Zhang2012, Lehmer2014, Peacock2016}). Fig.~\ref{fig:logN_logL_SM_All} shows stellar-mass normalized cumulative luminosity count distributions of each member galaxy in our samples. The galaxy sample used for stellar-mass normalized XLFs is partly different from that for SFR normalized XLFs because the samples should have stellar-mass information.

The overall shape of cumulative luminosity distribution is similar to each other. However, there are several outliers. The galaxy having the highest count is the Garland galaxy. Garland is a member galaxy of the M~81 group \citep{Karachentsev2002}. It is an irregular galaxy having a stellar mass of $3.7\times10^{6}\ {\rm M_\odot}$. Their origin and property are still under debate. The lowest count is NGC~0315, an elliptical galaxy, which has a stellar mass of $9.1\times10^{11}\ {\rm M_\odot}$. NGC~0315 is known as a low-ionization nuclear emission-line region AGN having an extended radio jet \citep{Healey2007, NGC0315}.

\begin{figure}
 \begin{center}
  \includegraphics[width=\linewidth]{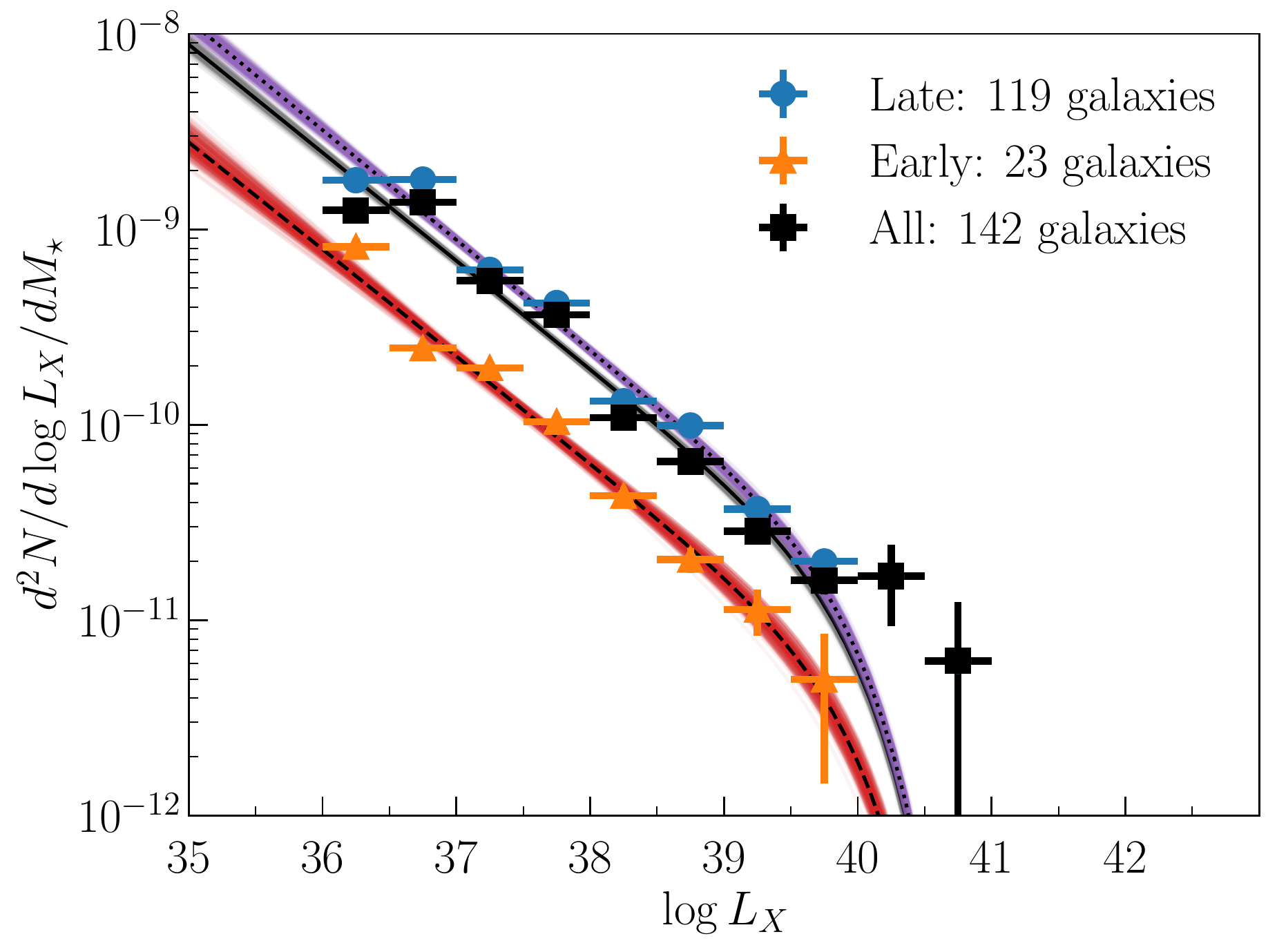}
 \end{center}
\caption{Same as Fig.~\ref{fig:dN_dlogL_Type}, but for stellar mass normalized XLFs.}\label{fig:dN_dlogL_SM_Type} 
\end{figure}

Same as Fig.~\ref{fig:dN_dlogL_Type}, Fig.~\ref{fig:dN_dlogL_SM_Type} shows the stellar mass normalized XLFs (${d^2N}/{d\log L_XdM_\star}$). In this plot, the maximum luminosity bin is lower than that in Fig.~\ref{fig:dN_dlogL_Type}. This is because we utilize the stellar mass available galaxies only. We fit XLFs using the same function as Eq.~\ref{eq:XLF} in the same manner, but replacing $\dot{\rho}_\star$ with $M_\star$. The fitting results are shown in Table.~\ref{tab:XLF} and Fig.~\ref{fig:dN_dlogL_SM_Type}. 

\begin{figure*}
 \begin{center}
  \includegraphics[width=0.495\linewidth]{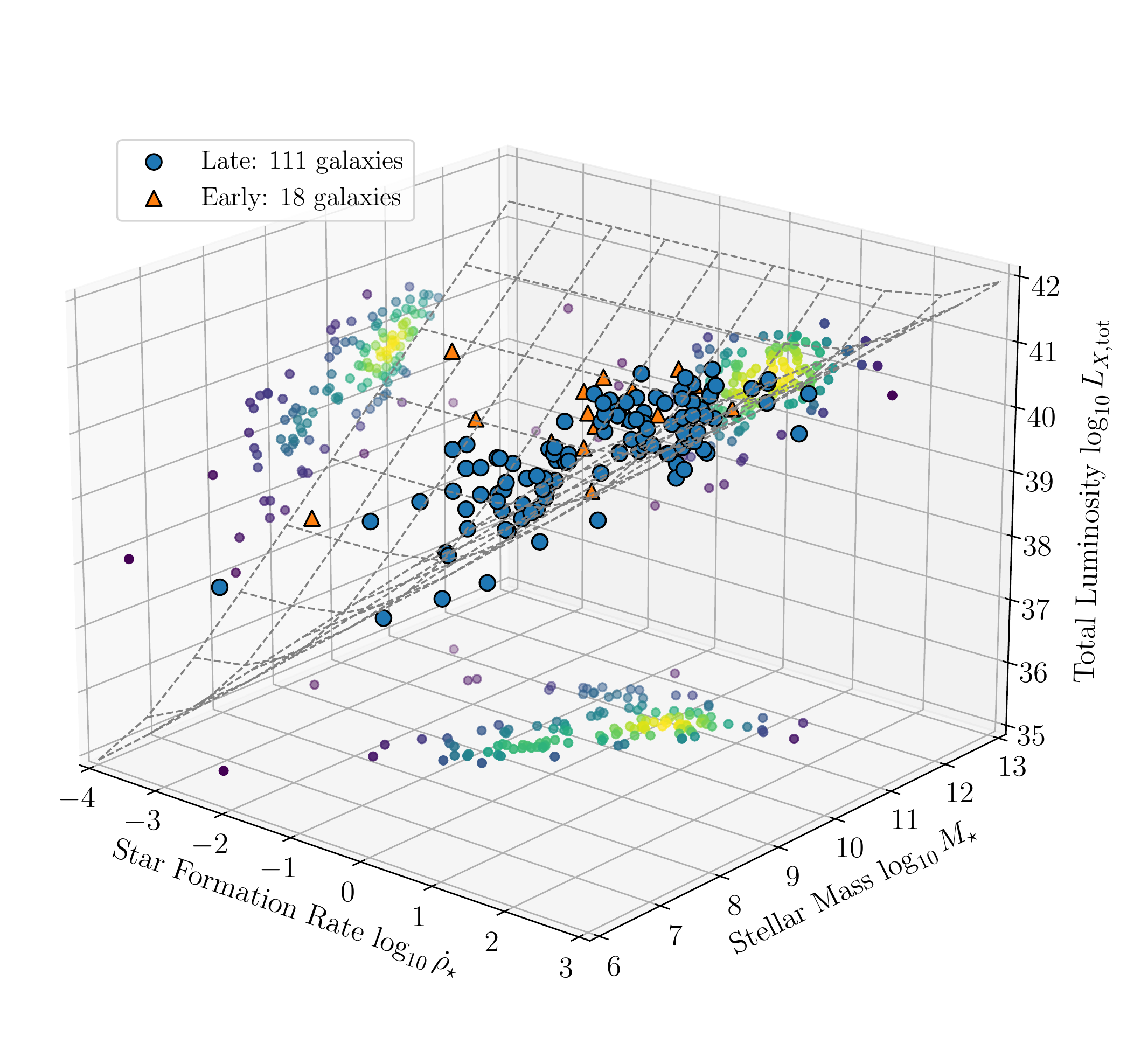}
  \includegraphics[width=0.495\linewidth]{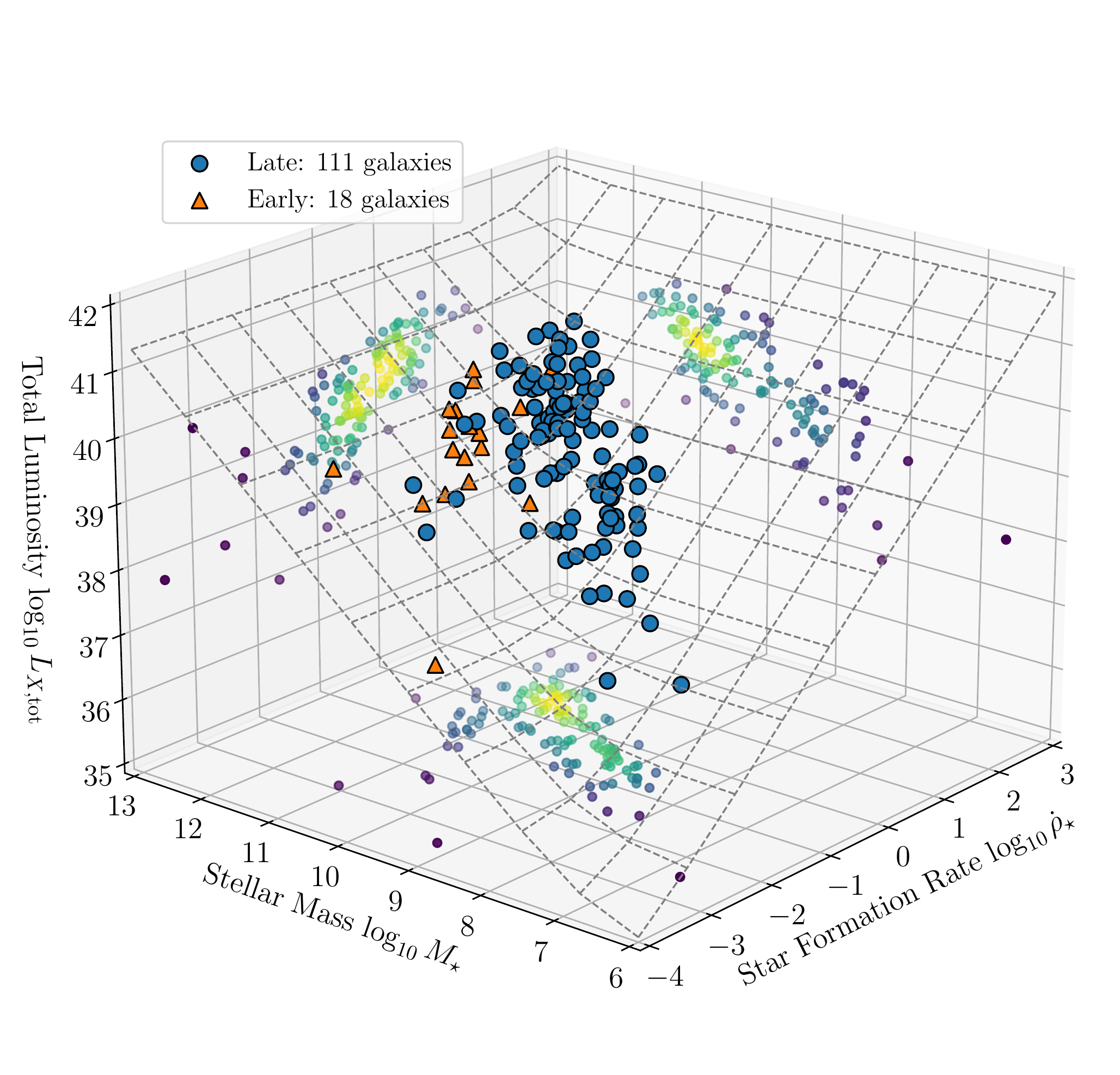}
 \end{center}
\caption{{\it Left}: Three-dimensional plot of the relation among $L_{\rm tot}$, $\dot{\rho}_{\star}$, $M_\star$. The circle and triangle data points correspond to late-type and early-type galaxies. Numbers of galaxy samples are indicated in the panel. The 2D projection is also shown in the figure. The dashed plane shows the fundamental plane of XRBs. {\it Right}: The same as in the {\it Left} panel, but from a different angle.}\label{fig:FP_3D} 
\end{figure*}

\subsection{Fundamental Plane of XRB hosting Galaxies}
\label{sec:FP}
SFR and stellar mass normalized XLFs are motivated by the fact that SFR traces the young stellar population, i.e., HMXBs, and stellar mass traces the old stellar population, i.e., LMXBs. However, as we see in our Galaxy, XRB populations in a galaxy are expected to be a mixture of HMXBs and LMXBs. Thus, it is naturally expected that both $\dot{\rho}_\star$ and $M_\star$ are important to characterize the XRB population in a galaxy. In this section, we investigate the relation among the integrated luminosity of XRBs $L_{X, \rm tot}$, $\dot{\rho}_\star$, and $M_\star$ of each galaxy. 

\citet{Lehmer2010ApJ...724..559L} pointed out that 2--10~keV X-ray emission of nearby galaxies correlates with both $\dot{\rho}_\star$ and $M_\star$ using 17 luminous infrared galaxies (LIRGs). \citet{Lehmer2016ApJ...825....7L} further investigated the redshift evolution of that correlation. In this paper, we compile {237} XRB host galaxies. Among them, {129} galaxies have both $\dot{\rho}_\star$ and $M_\star$ information. Since the previous studies integrated the whole X-ray emission of galaxies, there was about 10\% level of flux contamination by galactic hot gas emission ($\sim0.5$--$1$~keV). By extracting individual XRBs, we can directly estimate the total X-ray emission from XRBs.

Fig.~\ref{fig:FP_3D} shows the three-dimensional plot of the relation among $L_{X, \rm tot}$, $\dot{\rho}_{\star}$, and $M_\star$. $L_{X, \rm tot}$ is the summation of X-ray luminosities of XRBs with $\log L_X\ge36$ in a galaxy. These two panels show the same data but from different viewing angles. As it is clearly seen from the plots, higher $\dot{\rho}_\star$ and higher $M_\star$ tend to have higher $L_{X, \rm tot}$. This is because there are three relations: the main-sequence relation between $\dot{\rho}_\star$ and $M_\star$ (e.g., \cite{Noeske2007,Salim2007}), the correlation between  $\dot{\rho}_\star$ and $L_{X, \rm tot}$ (e.g., \cite{Grimm2003,Mineo2012}), and the correlation between $M_\star$ and $L_{X, \rm tot}$ (e.g., \cite{Gilfanov2004}). The Spearman correlation coefficients of those relations are {$\rho_S=0.65$, $0.58$, and $0.53$}, respectively, indicating the existence of positive correlations among those parameters.

\begin{figure}
 \begin{center}
  \includegraphics[width=\linewidth]{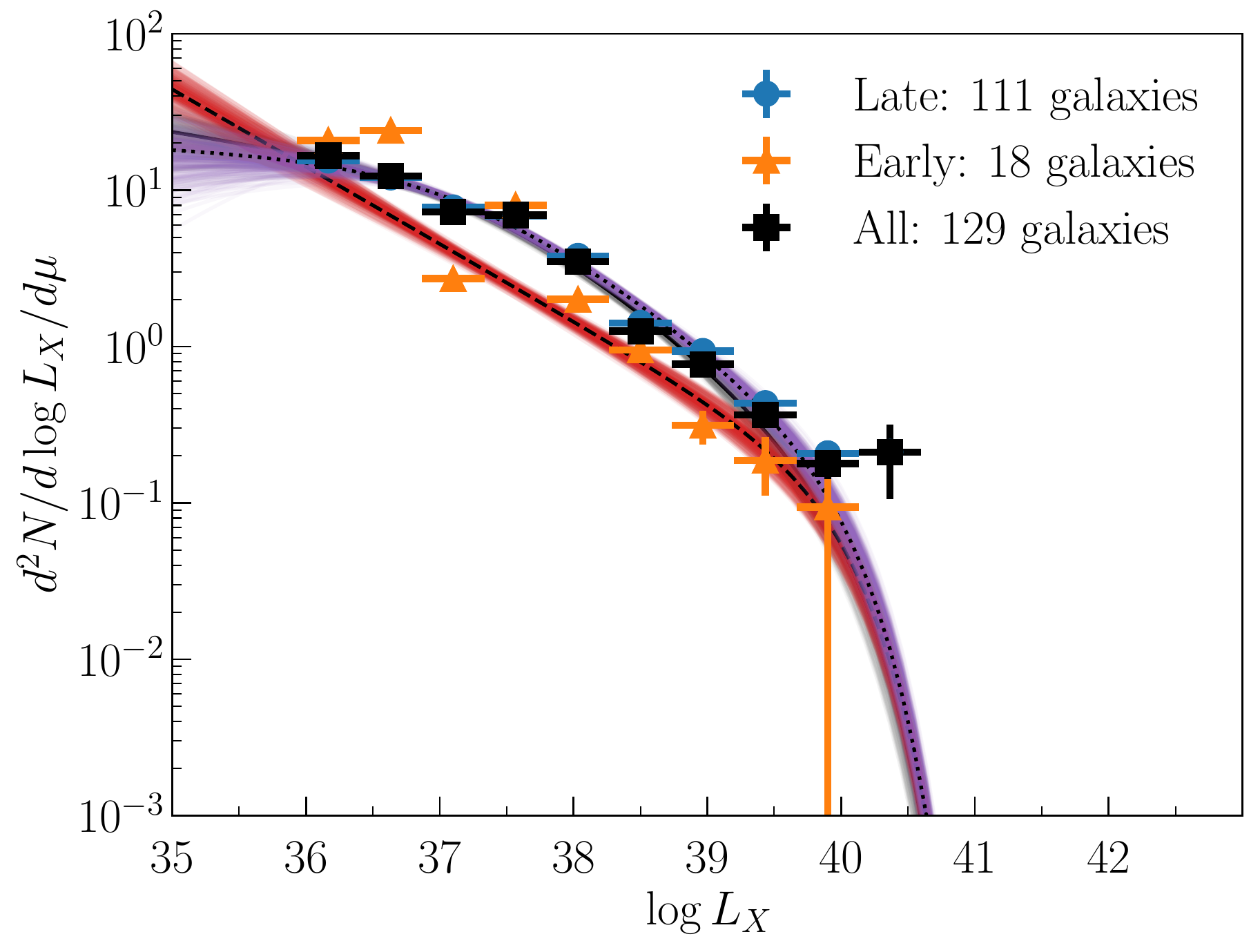}
 \end{center}
\caption{Same as Fig.~\ref{fig:dN_dlogL_Type}, but normalized by $\mu$.}\label{fig:dN_dlogL_Mu} 
\end{figure}

\begin{figure*}
 \begin{center}
  \includegraphics[width=0.85\linewidth]{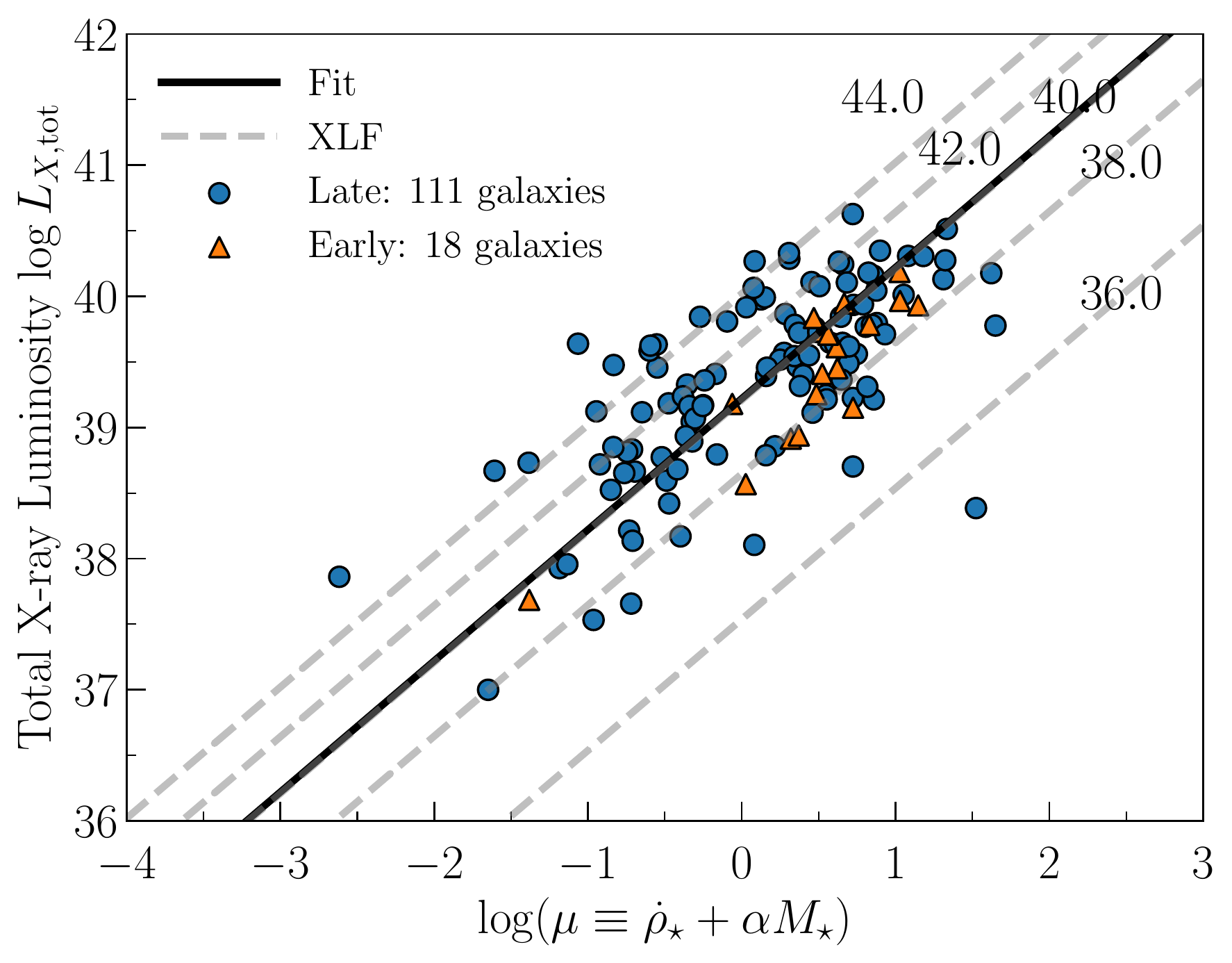}
 \end{center}
\caption{Comparison between $L_{X, \rm tot}$ and $\mu\equiv \dot{\rho}_\star + \alpha M_\star$ of XRB hosting galaxies, where $\alpha={3.36\times10^{-11}}~{\rm yr}^{-1}$. Circle and triangle data points correspond to late-type and early-type galaxies, respectively. The solid line shows the linear regression fit. The dashed lines show the expected relation from $\mu$-normalized XLFs with different $\log L_c$ as indicated in the figure. The dashed line with $\log L_c={40.0}$ overlaps with the fitting result.}\label{fig:Ltot_XRB} 
\end{figure*}

Considering the relation among these three parameters, here, we propose to introduce a new parameter defined as: 
\begin{equation}
\mu(\dot{\rho}_\star, M_\star)\equiv \dot\rho_\star	 + \alpha M_\star.
\label{eq:mu_def}
\end{equation}
The first term of the right-handed side is the star formation rate which represents the young stellar population (i.e., HMXBs), while the second term is linearly related to the stellar mass which represents the old stellar population (i.e., LMXBs). $\alpha$ controls the ratio between these two populations in the unit of ${\rm yr}^{-1}$. The physical meaning of $\alpha$ is discussed in \S.~\ref{sec:SXBPS}. Then, from Fig.~\ref{fig:FP_3D}, we have

\begin{equation}
\log L_{X, \rm tot} = ({38.80^{+0.09}_{-0.12}}) + \log \mu(\dot{\rho}_\star, M_\star),
\label{eq:Ltot_mu}
\end{equation}
with
\begin{equation}
\alpha = {(3.36\pm1.40)\times10^{-11}}\ {\rm yr^{-1}}.
\label{eq:alpha}
\end{equation}
The corresponding fundamental area is shown in Fig.~\ref{fig:FP_3D}. By introducing $\mu$, we can reproduce both $L_{X, \rm tot}$ and $\dot{\rho}_\star$ and $L_{X, \rm tot}$ and $M_\star$ relations known in XRB hosting galaxies simultaneously. Here, utilizing nearby 17 LIRGs, \citet{Lehmer2010ApJ...724..559L} reported as {$\log L_{X, \rm tot} = (39.22^{+0.05}_{-0.06}) + \log \mu(\dot{\rho}_\star, M_\star),$
with $\alpha = (5.45\pm0.76)\times10^{-11}\ {\rm yr^{-1}},$}
which is consistent with our result. We note that \citet{Lehmer2010ApJ...724..559L} adopted the Kroupa IMF \citep{Kroupa2001MNRAS.322..231K}.

Utilizing $\mu$, we can derive the $\mu$ normalized XLFs (${d^2N}/{d\log L_Xd\mu}$). The total number of galaxies is {129} which have both $\dot{\rho}_\star$ and $M_\star$. We fit XLFs using the same function as Eq.~\ref{eq:XLF} in the same manner, but replacing $\dot{\rho}_\star$ with $\mu$. The fitting results are shown in Table.~\ref{tab:XLF} and Fig.~\ref{fig:dN_dlogL_Mu}. We fix $\log L_c={40.0}$. By introducing $\mu$, the XLFs of the late, early, and all type galaxies show similar shapes indicating a universal trend in the local Universe. The similarity of XLF shapes indicates a single component dominates the XLFs, even though it is a mixture of HMXBs and LMXBs. This suggests that HMXBs and LMXBs may have a similar XLF shape.

Fig.~\ref{fig:Ltot_XRB} shows the relation between $\mu$ and $L_{X, {\rm tot}}$, corresponding to the fundamental plane relation. These two parameters show a positive correlation. The Spearman rank correlation coefficient is {0.68, 0.78, and 0.67} for late, early, and all type galaxies, respectively, with $p$-value of $\ll10^{-5}$. A linear regression line corresponding to Eq.~\ref{eq:Ltot_mu} is also shown as the solid line in the Fig.~\ref{fig:Ltot_XRB}. We note that we can fit the data with higher-order equations. However, as discussed in \S.\ref{sec:SFRXLF} and \S.\ref{sec:SMXLF}, the definition of Eqs.~\ref{eq:mu_def} and \ref{eq:Ltot_mu} is a physically reasonable choice.

We can reproduce this relation from the $\mu$ normalized XLF. We show the integrated luminosity for various $\mu$ in Fig.~\ref{fig:Ltot_XRB} by the dashed line. The minimum and maximum luminosity is set to be $10^{36}$ and $10^{45}\  {\rm erg \  s^{-1}}$, respectively. Because of the existence of $L_c$, the choice of the maximum luminosity does not affect the results as far as we set it higher than $L_c$. Different dashed lines correspond to the different $L_c$ values. As presented in Fig.~\ref{fig:Ltot_XRB}, $\log L_c={40.0}$ almost matches with the regression line. Therefore, we set $\log L_c={40.0}$ in this paper. 


\section{Phenomenological XRB Population Synthesis Model}
\label{sec:SXBPS}

The $\mu$ normalized XLF reveals the bright-end XLF slope is $\gamma_2\sim0.6$ for any galaxies. It is interesting to see whether this $\gamma_2$ can be reproduced by considering a binary population synthesis model. Today, various binary population synthesis models are available in the literature (e.g., \cite{Belczynski2008, Belczynski2010, Belczynski2016_Nat, Belczynski2016,  Mandel2016, Pavlovskii2017, Marchant2017, Kruckow2018, Mapelli2018}). However, each model contains various detailed physical processes, and it is not easy to compare with our current data directly. Therefore, we construct a phenomenological XRB population synthesis model utilizing the latest stellar evolution model. 

First, we consider the HMXB population. Once we define the star formation rate $\dot{\rho_\star}$ and the IMF $dN/dM_\ZAMS$ shape at zero-age main sequence (ZAMS) phase in a galaxy, the stellar mass distribution formed in a unit time can be evaluated. $M_\ZAMS$ is the mass of ZAMS stars. Suppose we have a power-law IMF such as the Salpeter IMF, $dN/dM_\ZAMS\propto M_\ZAMS^{-\alpha_{\rm IMF}}$, the stellar mass function in a unit time is given as
\begin{equation}
\frac{d^2N}{dM_\ZAMS dt}= \frac{(2-\alpha_{\rm IMF})\dot{\rho}_\star}{M_{\ZAMS, \rm max}^{2-\alpha_{\rm IMF}} - M_{\ZAMS, \rm min}^{2-\alpha_{\rm IMF}}}M_\ZAMS^{-\alpha_{\rm IMF}},
	\label{eq:HMXB_zams}
\end{equation}
where $M_{\ZAMS, \rm max}$ and $M_{\ZAMS, \rm min}$ are the maximum and minimum mass of ZAMS stars.

We utilize a model by \citet{Spera2015} for the mass spectrum of compact remnants (neutron stars and black holes) after the stellar evolution, including the metallicity dependence (See Fig.~6 and Appendix C in \cite{Spera2015}). Here, the metal environment affects the resulting compact object mass distribution (e.g., \cite{Spera2015}). It is also known that metallicity of galaxies depends on $M_\star$ and $\dot{\rho}_\star$ \citep{Mannucci2010MNRAS.408.2115M, Yabe2012PASJ...64...60Y, Andrews2013ApJ...765..140A}. In this paper, since we consider the local Universe only, we fix the metallicity to the solar value.

Then, we have the compact object formation rate as
\begin{equation}
	\frac{d^2N}{dM_\co dt} = \frac{d^2N}{dM_\ZAMS dt}\frac{dM_\ZAMS}{dM_\co},
\end{equation}
{where $M_\co$ is the mass of the compact remnant.}
Following \citet{Spera2015}, we treat a compact remnant as a white dwarf when its final core mass is less than the Chandrasekhar mass ($1.4M_\odot$). The remnants with masses of $1.4M_\odot\le M_\co < 3.0 M_\odot$  are treated as neutron stars, while those with $M_\co\ge 3.0 M_\odot$ as black holes. 

Among produced compact remnants, we require binary systems in order to have accretion disk activity. We describe {the binary} fraction as $f_b$. Also, we require the fraction of X-ray emitting binaries among those binaries $f_\HMXB$  (i.e., in the mass transfer phase with an adequate mass accretion rate). Then, the mass function of compact objects having the HMXB activity can be approximated as 

\begin{equation}
	\frac{dN_\HMXB}{dM_{\co}} \approx f_b f_\HMXB t_\HMXB \frac{d^2N}{dM_\co dt},
	\label{eq:HMXB_co}
\end{equation}
where $t_\HMXB$ is the duration of HMXB activity. Here, in our Galaxy, a large fraction of massive stars ($\gtrsim70$\%, \cite{Sana2012}) are members of binary systems since their birth. For simplicity, we set the binary fraction as $f_b\sim0.7$. And, although $t_\HMXB$ is not well constrained, it is expected to be about 0.1~Myr \citep{Mineo2012}.

A study of Galactic accreting stellar-mass black holes indicates that the logarithmic Eddington ratio distribution is given by a normal distribution \citep{Reynolds2013,Finke2017},
\begin{equation}
	P_{\ell}(\log\ellE) = \frac{1}{\sqrt{2\pi}\sigma_{\ell}}
\exp\left\{ \frac{-(\log\ellE-\mu_{\ell})^2}{2\sigma_{\ell}^2}\right\},
\end{equation}
where $\ellE$ is the Eddington ratio. $\mu_{\ell}$ is $\sim-2$ and $\sigma_{\ell}\sim1$ for the Galactic XRBs including both HMXBs and LMXBs at all epochs \citep{Reynolds2013}.

The resulting XLF of HMXBs at an X-ray luminosity of $L_X$ can be evaluated as
\begin{eqnarray}
	\frac{dN_\HMXB}{dL_X} &=& \int d\log\ellE\frac{dN_\HMXB}{dM_\co}\frac{dM_\co}{dL_X} P_{\ell},
	\\ \nonumber
	&=& f_b f_\HMXB t_\HMXB \dot{\rho}_\star \int d\log\ellE\frac{dM_\co}{dL_X} P_{\ell}\\
	&\times&\frac{(2-\alpha_{\rm IMF})M_\ZAMS^{-\alpha_{\rm IMF}}}{M_{\ZAMS, \rm max}^{2-\alpha_{\rm IMF}} - M_{\ZAMS, \rm min}^{2-\alpha_{\rm IMF}}}\frac{dM_\ZAMS}{dM_\co}\label{eq:HMXB_XLF}
\end{eqnarray}
where we set $L_X = \ellE \LEdd(M_\co)$. $\LEdd$ is the Eddington luminosity $\approx1.26\times10^{38} (M_\co/M_\odot)\ {\rm erg\ s^{-1}}$.

Next, we consider the LMXB population. We can adopt the same method as in the HMXB population, but LMXBs are expected to follow the stellar mass. Therefore, we rewrite the equations \ref{eq:HMXB_zams} and \ref{eq:HMXB_co} which are  related to star formation as 
\begin{eqnarray}
\frac{dN}{dM_\ZAMS}&=& \frac{(2-\alpha_{\rm IMF})M_\star}{M_{\ZAMS, \rm max}^{2-\alpha_{\rm IMF}} - M_{\ZAMS, \rm min}^{2-\alpha_{\rm IMF}}}M_\ZAMS^{-\alpha_{\rm IMF}},
	\label{eq:LMXB_zams}\\
	\frac{dN_\LMXB}{dM_\co} &\approx& f_b f_\LMXB \frac{t_\LMXB}{t_{\galaxy}} \frac{dN}{dM_\co},
	\label{eq:LMXB_co}
\end{eqnarray}
where $f_\LMXB$ is the LMXB fraction among binaries having compact objects and $t_\LMXB$ is the duration of LMXB activity, which is expected to be typically $10$~Myr. $t_{\galaxy}$ is the age of the galaxy and we assume a constant star formation activity. By adopting Eq.~\ref{eq:HMXB_XLF}, we can evaluate $dN_\LMXB/dL_X$ as
\begin{eqnarray} \nonumber
	\frac{dN_\LMXB}{dL_X} &=& f_b f_\LMXB \frac{t_\LMXB}{t_{\galaxy}} M_\star \int d\log\ellE\frac{dM_\co}{dL_X} P_{\ell}\\
	&\times&\frac{(2-\alpha_{\rm IMF})M_\ZAMS^{-\alpha_{\rm IMF}}}{M_{\ZAMS, \rm max}^{2-\alpha_{\rm IMF}} - M_{\ZAMS, \rm min}^{2-\alpha_{\rm IMF}}}\frac{dM_\ZAMS}{dM_\co}.
	\label{eq:LXMB_XLF}
\end{eqnarray}
Here, both integral terms in the HMXB and LMXB XLFs are the same, which are described as $\phi_X(L_X)$ hereinafter.

\begin{figure}
 \begin{center}
  \includegraphics[width=\linewidth]{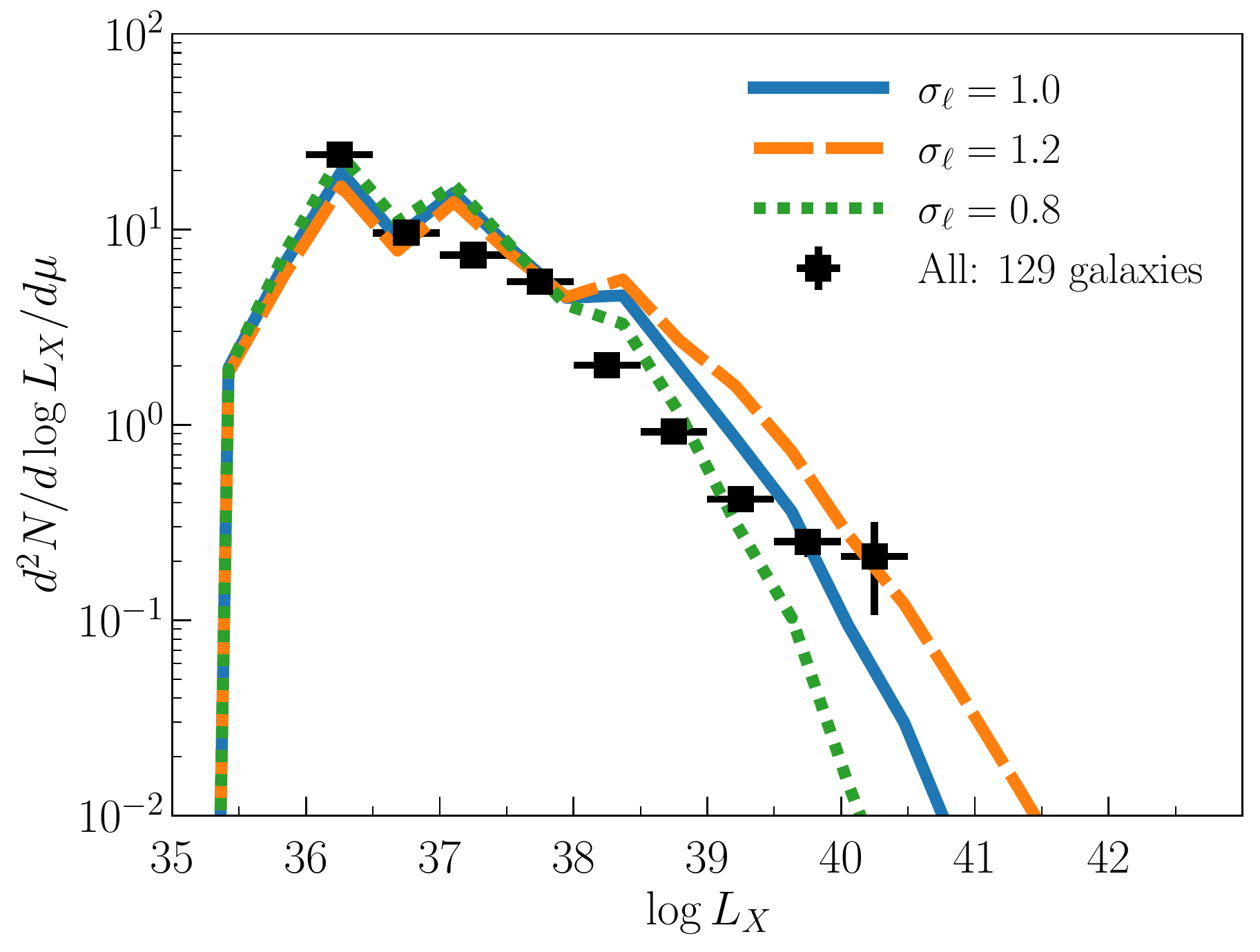}
 \end{center}
\caption{The $\mu$ normalized XRB XLF based on the phenomenological binary population synthesis model (PBPS). The data points are for full galaxy sample presented in Fig.~\ref{fig:dN_dlogL_Mu}. The solid, dashed, and dotted curves show the SBPS model curve with different $\sigma$ as indicated in the figure. The metallicity is fixed to $Z=0.01$.}\label{fig:dN_dL_SBPS} 
\end{figure}

Then, the whole XRB XLF becomes
\begin{equation}
\frac{dN}{dL_X} = \frac{dN_\HMXB}{dL_X} + \frac{dN_\LMXB}{dL_X}.
\end{equation}
As $dN_\HMXB/dL_X\propto \dot{\rho}_\star$ and $dN_\LMXB/dL_X\propto M_\star$ and by adopting Eqs.~\ref{eq:HMXB_XLF} and\ref{eq:LXMB_XLF},
\begin{equation}
	\frac{dN}{dL_X} = \left(f_\HMXB t_\HMXB \dot{\rho}_\star + f_\LMXB \frac{t_\LMXB}{t_{\galaxy}} M_\star \right) f_b\phi_X(L_X).
\end{equation}
By comparing to the $\mu$ normalized XLF {(See Eq.~\ref{eq:mu_def})}, we should have
\begin{equation}
	\alpha = \frac{f_\LMXB t_\LMXB}{f_\HMXB t_\HMXB} \frac{1}{t_\galaxy}.
	\label{eq:alpha_sbps}
\end{equation}
The normalization is determined by $f_b f_\HMXB t_\HMXB$. {Therefore, $\alpha$ represents the ratio between \HMXB\ and \LMXB\ in a galaxy lifetime.}

\begin{figure*}
 \begin{center}
  \includegraphics[width=0.495\linewidth]{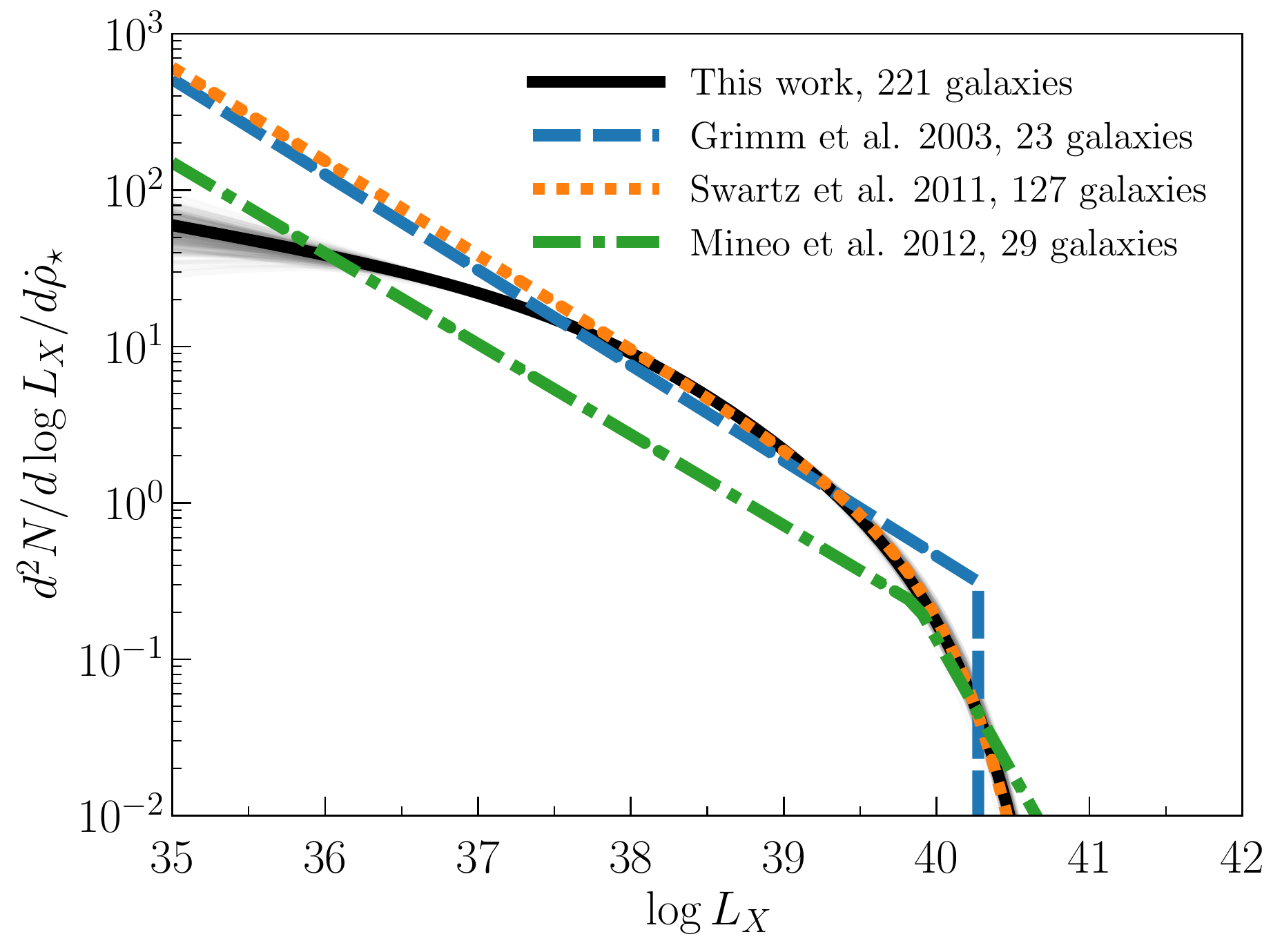}
  \includegraphics[width=0.495\linewidth]{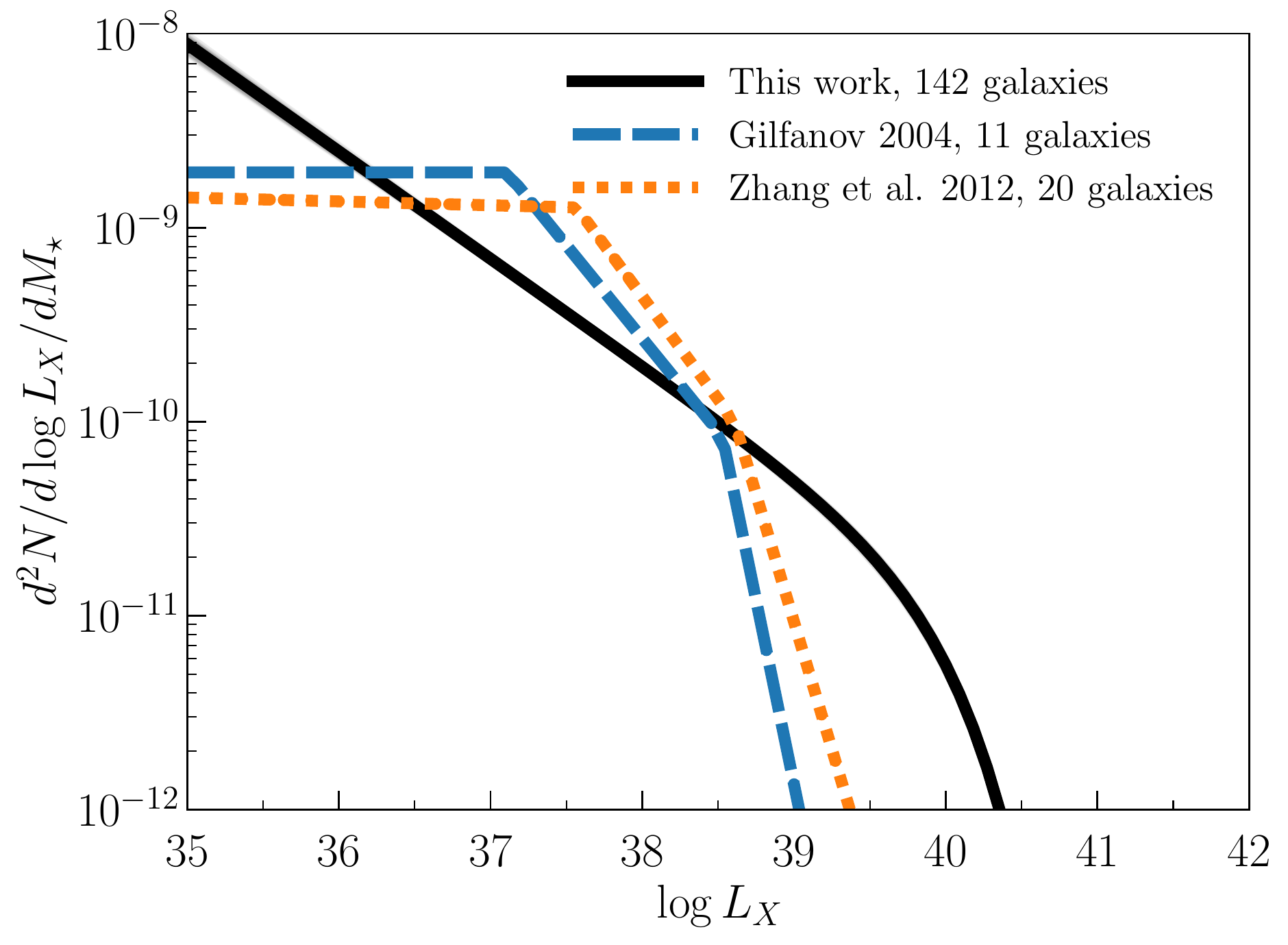}
 \end{center}
\caption{{\it Left}: $\dot{\rho}_\star$ normalized XLFs. Solid, dashed, dotted, dot-dashed line corresponds to our study, \citet{Grimm2003}, \citet{Swartz2011}, and \citet{Mineo2012}, respectively. The number of sampled galaxies is also shown in the panel. {\it Right}: The same as in the {\it Left} panel, but for $M_\star$ normalized XLFs. Solid, dashed, and dotted line corresponds to our study, \citet{Gilfanov2004}, and \citet{Zhang2012}, respectively.}\label{fig:XLF_comp} 
\end{figure*}

Fig.~\ref{fig:dN_dL_SBPS} shows the $\mu$ normalized XRB XLF based on our phenomenological binary population synthesis (PBPS) model described above {with various $\sigma_\ell$}. The binned XRB XLF is also shown. The model with $\sigma_{\ell}={1.0}$ well reproduces the data which is in the consistent range with the Galactic XRB distribution \citep{Reynolds2013}. For the other parameters except for normalization, we do not change. {This result tells that the XLF slopes are determined by the IMF and the Eddington ratio distribution.} 

In order to match the data, we require the fraction of HMXB in compact binary systems formed in a certain time as
\begin{equation}
	f_\HMXB \approx 0.09 \left(\frac{t_\HMXB}{0.1~{\rm Myr}}\right)^{-1}\left(\frac{f_b}{0.7}\right)^{-1}.
\end{equation}
This manifests that $t_\HMXB$ can not be $\ll 0.01$~Myr otherwise $f_\HMXB>1$. As described above, $t_\HMXB$ is typically about 0.1~Myr \citep{Mineo2012}. From binary evolution calculations, $t_\HMXB\sim0.01$~Myr for supergiant systems and $t_\HMXB\sim0.1$~Myr for Be/X binaries. About 60\% of known high-mass X-ray binaries are Be/X systems, while $\sim30$\% are supergiant systems (e.g., \cite{Liu2006}).

	From Eq.~\ref{eq:alpha_sbps}, the fraction of LMXB in compact binary systems formed in a certain time is
\begin{eqnarray}\nonumber
	f_\LMXB &\approx& {4.4}\times10^{-4} \left(\frac{\alpha}{{3.36}\times10^{-11}~{\rm yr^{-1}}}\right)\\
	&\times&\left(\frac{t_\galaxy}{10~{\rm Gyr}}\right)\left(\frac{t_\LMXB}{10~{\rm Myr}}\right)^{-1}\left(\frac{f_b}{0.7}\right)^{-1}.
\end{eqnarray}
Thus, LMXBs are rare objects per star formation activity, but we can observe them a lot because of its long lifetime comparing to HMXBs. The reason for the difference between $f_\HMXB$ and $f_\LMXB$ may be due to the required mass ratio. In our own Galaxy, massive stars are known to be members of binary systems whose mass ratio distribution is flat \citep{Kobulnicky2007,Sana2012,Kobulnicky2014}. Here, LMXBs requires a high mass ratio because the donor star is a low-mass star (e.g., \cite{Kalogera1998,Tauris2006}). Therefore, it is natural to expect a low $f_\LMXB$ value. This equation also indicates that younger galaxies will have a lesser LMXB population.


\section{Discussion}
\label{sec:discussion}

\subsection{Comparison with Previous Studies of XRB XLFs}
The $\dot{\rho}_\star$ and $M_\star$ normalized XRB XLFs have been studies in literature. In this section, we compare our results with the previous studies on those XLFs.

The left panel of Fig.~\ref{fig:XLF_comp} shows the $\dot{\rho}_\star$ normalized XRB XLFs of our study, \citet{Grimm2003}, \citet{Swartz2011}, and \citet{Mineo2012}. We note that \citet{Swartz2011} used ULX data only. As seen from the figure, the slope of the XLFs are mostly consistent in $38\lesssim \log L_X \lesssim 40$, while the normalization of \citet{Mineo2012} is slightly smaller than the other works. One remarkable difference is seen in the low-luminosity regime. Even though the model of \citet{Swartz2011} is just an extrapolation, the \citet{Grimm2003} and \citet{Mineo2012} covered the luminosity range down to $\log L_X\sim35$. The reason for the lower estimates of our study is likely due to the incompleteness of the observations. In  \citet{Grimm2003} and \citet{Mineo2012}, the incompleteness of each observation is corrected, but we did not correct it as explained in \S~\ref{sec:SFRXLF}. However, we note that, as seen in Fig.~\ref{fig:dN_dL_SBPS}, the PBPS model shows a drop in XLF at lower luminosity due to the Eddington ratio and mass distribution.

The right panel of Fig.~\ref{fig:XLF_comp} shows the $M_\star$ normalized XRB XLF of our study, \citet{Gilfanov2004}, and \citet{Zhang2012}. As seen from the figure, the shape and normalization of the XLFs are mostly consistent among these three works, but our studies show the existence of luminous XRBs in galaxies. This may be due to our large galaxy sample, roughly 10 times more than those in the previous works. As the main parent galaxy sample class was early-type galaxy in \citet{Gilfanov2004} and \citet{Zhang2012}, their XLFs indicate fewer ULXs in early-type galaxies. However, it is known that elliptical galaxies host numerous ULXs \citep{Swartz2004}, which is qualitatively consistent with our result. Although the observational incompleteness is corrected for \citet{Gilfanov2004} and \citet{Zhang2012}, the faint-end slope is similar among these three works. This is because the correction effect of the observational incompleteness is not significant in these studies (see Table. 4 in \cite{Zhang2012}).

\subsection{Galaxy Sample Selection}
Some of our galaxy samples do not have information of either of $\dot{\rho}_\star$ or $M_\star$. Here, in our sample, the SFR completeness is about 93\%, which indicates biases on SFR information is not significant. However, the $\mu$ completeness is about {54}\%, indicating that our $\mu$ normalized XLF results can be affected by the selection effect. 

The observational status of each galaxy differs one by one. When SFR or stellar mass information is unavailable, it is not easy to get further information. Some of them may not even have measurements. Therefore, it is not straightforward to understand the selection effect. However, we can at least test whether the $\mu$ selected samples differ from the full sample using the $\dot{\rho}_\star$ normalized XLFs.

\begin{figure}
 \begin{center}
  \includegraphics[width=\linewidth]{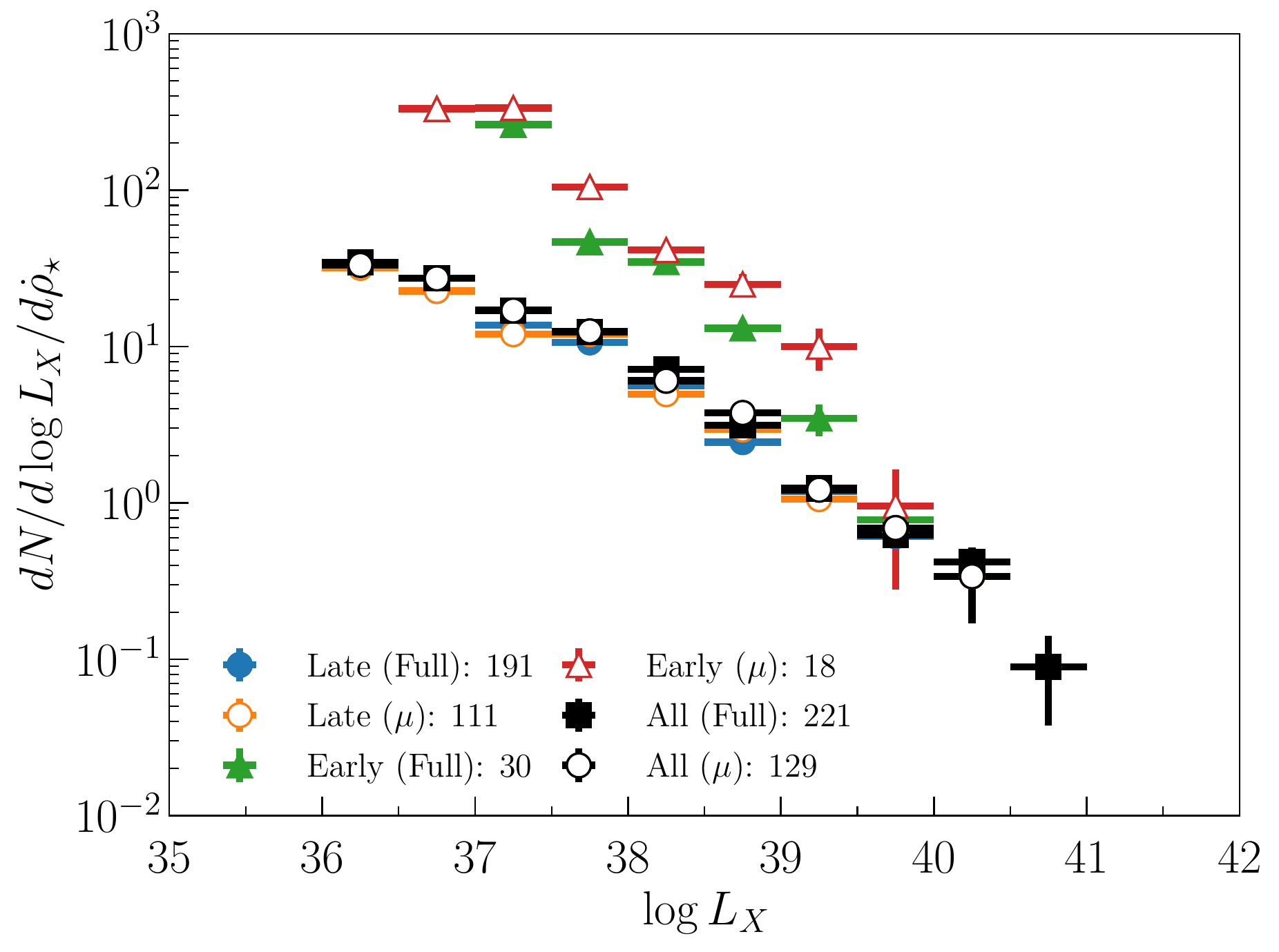}
 \end{center}
\caption{Comparison of the star formation rate normalized XLF of XRBs in different parent galaxy samples. Filled points correspond all the SFR available galaxies, while open symbols are restricted to the $\mu$ available samples. }\label{fig:dN_dlogL_Sample} 
\end{figure}

Fig.~\ref{fig:dN_dlogL_Sample} shows the comparison of the $\dot{\rho}_\star$ normalized XLFs of the full galaxy sample and the $\mu$ selected samples. As both data overlap, we do not see any clear deviation between these two galaxy samples. Therefore, the selection effect on the $\mu$ selected samples is insignificant.

\subsection{Effect of Background AGNs}
{AGNs, dominating the X-ray sky, would contaminate our XRB samples. To remove those background AGNs, we subtract known AGNs using available AGN catalogs \citep{Veron2010, Assef2018} and select galaxies whose X-ray source density is higher than AGN surface density. There is a less biased method. In this method, we estimate the number of possible background AGNs at a given flux within an area of a galaxy using AGN flux distribution, then subtract them from observed XRB XLFs assuming all the AGNs at a distance of that galaxy. However, this method requires a correction of detailed position dependence of the detection efficiency. In this paper, which we do not take into account this, although we adopt the flux threshold where almost uniform detection efficiency achieved \citep{Evans2010}.}

\begin{figure}
 \begin{center}
  \includegraphics[width=\linewidth]{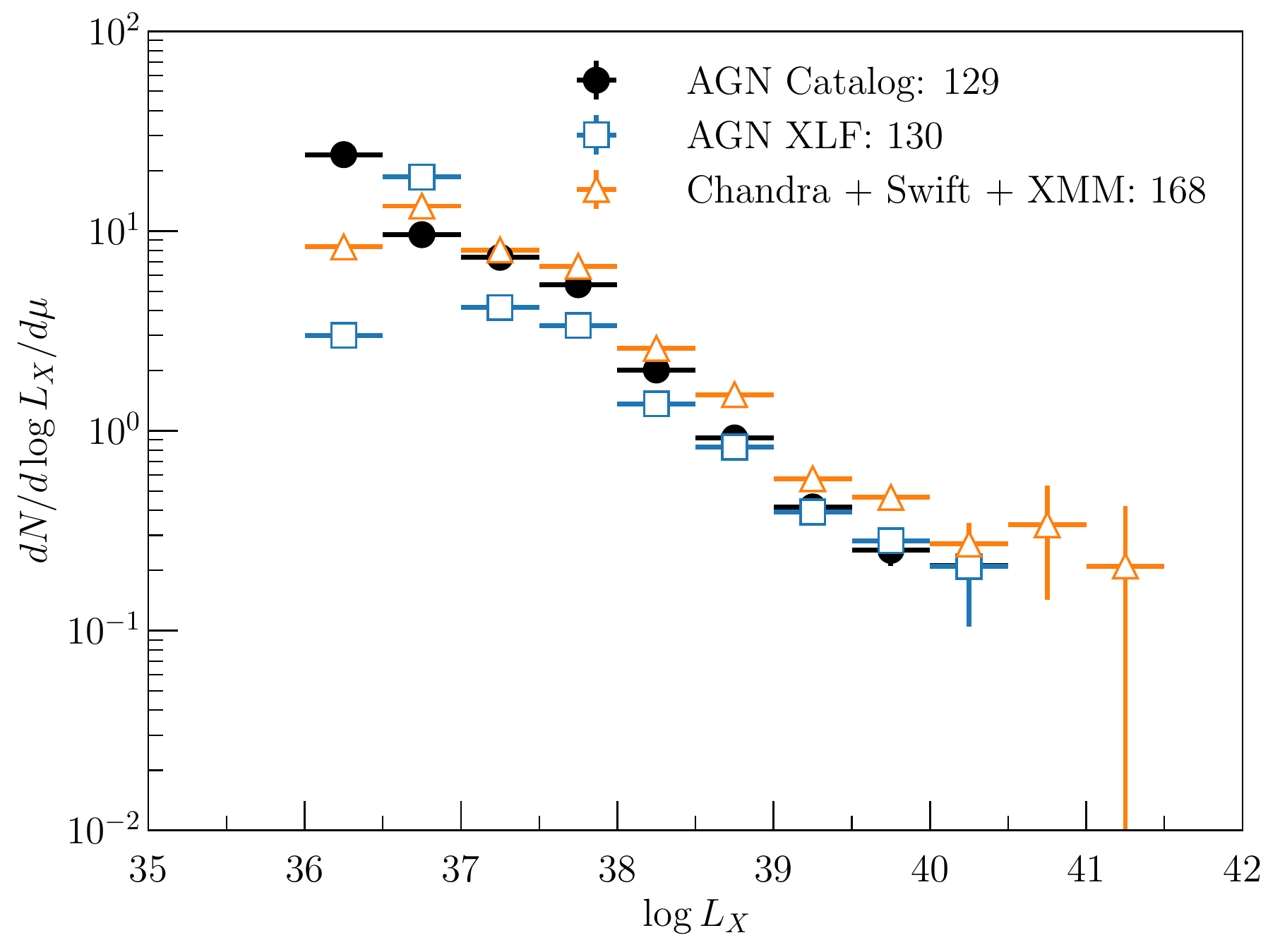}
 \end{center}
\caption{Comparison of the $\mu$ normalized XLF of XRBs among different catalogs. Filled data points correspond to our fiducial samples, while open square data points subtracts AGNs using AGN XLF \citep{Ueda2014} and open triangle data combines the Chandra-Swift-XMM catalogs.}\label{fig:dN_dlogL_AGN} 
\end{figure}

{To investigate the effect of possible background AGN contamination, we recalculate the $\mu$ normalized XLFs by subtracting background AGNs using AGN XLF \citep{Ueda2014}} {as described above, but without taking into account the detailed position dependence of the detection efficiency in each observation.} Fig.~\ref{fig:dN_dlogL_AGN} shows the comparison of recalculated XLF with our fiducial XLF. As shown in the figure, {although we see a clear deficit at the lowest luminosity band,} the two XLFs do not show clear difference at higher luminosity ranges. This {may indicate} that our AGN subtraction method would not significantly affect our main results. {We note that at lower-luminosity ranges, XLF density becomes lower for the AGN XLF method. This is because we do not take into account the detailed detection efficiency which generally drops at fainter fluxes.} 

{
\subsection{Swift and XMM-Newton Source Catalogs}
The Neil Gehrels {\it Swift} observatory and the {\it XMM-Newton} observatory also observe the X-ray. Both observatories recently released their source catalogs the 2SXPS catalog \citep{Evans2020} \footnote{\url{https://www.swift.ac.uk/2SXPS/}} and the 4XMM-DR9 catalog \citep{Webb2020} \footnote{\url{http://xmmssc.irap.omp.eu/Catalogue/4XMM-DR9/4XMM_DR9.html}}, respectively. The 2SXPS and 4XMM-DR9 catalog contains 206,335 and 550,124 unique X-ray sources in the sky, respectively. Although their angular resolution is not as good as that of {\it Chandra}, these samples help us to expand our XRB samples. We add these catalogs following the same selection criteria as \sout{done} for CSC2. However, the 2SXPS provides stacked data only.

There are significant amount of overlapped sources among the catalogs. To avoid such overlaps, we match the catalogs with each other and searched neighboring sources within a tolerance radius $r_{\rm tol}$ (see also \cite{Itoh2020}). For an X-ray source catalog, we collect all the X-ray sources from the other catalogs within 60~arcsec, which is large enough comparing to the positional uncertainty of the three X-ray satellites. To set the background contamination less than 5\%, we set the $r_{\rm tol}=3.3^{\prime\prime}$, $3.5^{\prime\prime}$, and $5.8^{\prime\prime}$ for the matching of {CSC2 - 4XMM-DR9},  {CSC2 - 2SXPS}, and {4XMM-DR9 - 2SXPS}, respectively. We prioritized X-ray sources in the order of CSC2, 4XMM-DR9, and 2SXPS, considering their angular resolutions. According to the prioritization, we remove objects matched in other catalogs.

By combining three X-ray catalog and matching with galaxy catalogs, then it includes {5757} X-ray sources associated with {311} galaxies. Among those XRBs, {635} are ULX candidates. We reconstruct the $\mu$ normalized XLF using this larger sample data set. {Fig~\ref{fig:dN_dlogL_AGN} also }shows the $\mu$ normalized XLFs based on the {\it Chandra}-{\it Swift}-{\it XMM} combined catalog and the {\it Chandra} only catalog. At $\log L_{X}<40.5$, {both distributions show} the similar structure, while the combined catalog extend the high-luminosity end toward $\log L_X=41.5$. This is because of four 2SXPS sources having $\log L_X > 40.5$. As described above, the 2SXPS catalog provide the stacked information, which may result in higher fluxes than typical fluxes. Further careful treatment of 2SXPS and XMM-DR9 catalogs including e-ROSITA will help to understand the detailed statistical properties of XRBs.}


\section{Conclusion}
\label{sec:conclusion}

In this paper, we construct a new catalog of XRBs in external galaxies utilizing the latest {{\it Chandra} \citep{Evans2010,Evans2020_CSC2} source catalog}. We match these X-ray source catalogs with two nearby galaxy catalogs: the LVG catalog and the {\it IRAS} catalog. Our catalog contains {4430} XRBs, including {378} ULXs,  associated with {237} galaxies. $\sim84$\% of those XRB hosting galaxies are late-type. The XRB host galaxies reproduce the so-called galaxy main-sequence relation seen in the local galaxies (e.g., \cite{Noeske2007,Salim2007}.

$\gtrsim 99$\% of our XRB samples are off-nucleus objects. {There is no apparent difference in the spatial distributions of XRBs and ULXs, while spatial distributions of XRBs between late and early-type galaxies show difference}.

As demonstrated in previous studies, we can reproduce the SFR ($\dot{\rho}_\star$) and stellar mass ($M_\star$) normalized XLF relations and both of them are well described by a broken-power-law XLF model. With our XRB samples, we further investigate a relation among the integrated luminosity of XRBs $L_{X, \rm tot}$, $\dot{\rho}_\star$, and $M_\star$ in each galaxy. We find that there is a fundamental plane in those three parameters as {$\log L_{X, \rm tot} = (38.80^{+0.09}_{-0.12}) + \log \mu$, where $\mu\equiv \dot\rho_\star	 + \alpha M_\star$. $\alpha$ is determined as $(3.36\pm1.40)\times10^{-11}\ {\rm yr^{-1}}$}. The $\mu$ normalized XLFs of late-type and early-type galaxies are almost equivalent, suggesting that HMXBs and LMXBs may have a similar XLF shape. Furthermore, the $\mu$ normalized XLF can reproduce the correlation between $\mu$ and $L_{X, {\rm tot}}$ of our XRB host galaxy samples.

In order to investigate the fundamental plane, we construct a phenomenological binary population synthesis model. We find that the high mass XRB (HMXB) and low mass XRB (LMXB) fraction in formed compact object binary systems is $\sim9$\% and {$0.04$}\%, respectively. Thus, LMXBs are rare objects per star formation activity, but we can observe many of them because of their long lifetime comparing to HMXBs.

{
To increase the number of potential XRBs, we further add the  2SXPS catalog \citep{Evans2020} and the 4XMM-DR9 catalog \citep{Webb2020} based on the {\it Swift} and {\it XMM-Newton} observations respectively. Our XRB catalog also provides these information. 

By adding these two X-ray catalogs, the sample size increases to 5757 XRBs associated within 311 galaxies. However, we note that {\it Chandra} has better angular resolution than the other two and that the 2SXPS catalog provides the stacked flux information only. Therefore, for the analysis of XLFs, it would be better to use the {\it Chandra} sources only.}

\begin{ack}
{We thank the anonymous referee for constructive comments and suggestions.} We would like to thank Shinya Nakashima and Takaaki Tanaka for useful discussions and comments. This research has made use of the NASA/IPAC Extragalactic Database, which is funded by the National Aeronautics and Space Administration and operated by the California Institute of Technology. This research has made use of the SIMBAD database, operated at CDS, Strasbourg, France. This research has made use of the SIMBAD database,
operated at CDS, Strasbourg, France. Y.I. is supported by JSPS KAKENHI Grant Number JP18H05458 and JP19K14772. K.Y. is supported by JSPS KAKENHI Grant Number JP18K13578. Y.U. is supported by JSPS KAKENHI Grant Number JP17H06362. This work was supported by World Premier International Research Center Initiative (WPI), MEXT, Japan.
\end{ack}


\end{document}